\documentclass[sn-mathphys]{sn-jnl}

\usepackage{graphicx}%
\usepackage{multirow}%
\usepackage{amsmath,amssymb,amsfonts}%
\usepackage{amsthm}%
\usepackage{mathrsfs}%
\usepackage[title]{appendix}%
\usepackage{xcolor}%
\usepackage{textcomp}%
\usepackage{manyfoot}%
\usepackage{booktabs}%
\usepackage{algpseudocode}%
\usepackage{listings}%
\usepackage{diagbox}

\theoremstyle{thmstyleone}%
\theoremstyle{thmstyletwo}%

\theoremstyle{thmstylethree}%

\usepackage{amssymb}
\usepackage{xspace}
\usepackage{amsmath}
\usepackage{multirow}
\usepackage{array}
\usepackage{booktabs}
\usepackage{cleveref}
\usepackage{balance}
\usepackage[linesnumbered, ruled, vlined]{algorithm2e}
\usepackage{subfigure}
\usepackage{graphicx}
\usepackage{epstopdf}

\crefname{equation}{Eq.}{Eqs.}
\crefname{table}{Table}{Tables}
\crefname{figure}{Fig.}{Figures}
\crefname{algorithm}{Algorithm}{Algorithms}
\crefname{algocf}{Algorithm}{Algorithms}
\crefname{section}{Sec.}{Secs.}

 \usepackage{amsthm}

\begin{document}
	
	\title[Article Title]{RCoCo: Contrastive Collective Link Prediction across Multiplex Network in Riemannian Space}
	
	
	\author*[1]{\fnm{Li} \sur{Sun}}\email{ccesunli@ncepu.edu.cn}
	
	\author[1]{\fnm{Mengjie} \sur{Li}}\email{mengjie@ncepu.edu.cn}
	
	\author*[2]{\fnm{Yong} \sur{Yang}}\email{yangyongxp@163.com}
	
	\author[3]{\fnm{Xiao} \sur{Li}}\email{lixiao@cnu.edu.cn}
	
	\author[4]{\fnm{Lin} \sur{Liu}}\email{liulin@besti.edu.cn}
	
	\author[5]{\fnm{Pengfei} \sur{Zhang}}\email{zpf@aust.edu.cn}
	
	\author[6]{\fnm{Haohua} \sur{Du}}\email{duhaohua@buaa.edu.cn}

	\affil*[1]{\orgname{North China Electric Power University}, \orgaddress{\postcode{102206}, \state{Beijing}, \country{China}}}
	
	\affil[2]{\orgname{State Grid Handan Electric Power Supply Company}, \orgaddress{\postcode{056000}, \city{Handan}, \state{Hebei}, \country{China}}}
	
	\affil[3]{\orgname{Capital Normal University}, \orgaddress{\postcode{100048}, \state{Beijing}, \country{China}}}
	
	\affil[4]{\orgname{Beijing Institute of Electronic Science and Technology}, \orgaddress{\postcode{100070}, \state{Beijing}, \country{China}}}
	
	\affil[5]{\orgname{Anhui University of Science and Technology}, \orgaddress{\postcode{232001}, \city{Huainan}, \country{China}}}
	
	\affil[6]{\orgname{Beihang University}, \orgaddress{\postcode{100191}, \state{Beijing}, \country{China}}}

	

\abstract{
Link prediction typically studies the probability of future interconnection among nodes with the observation in a single social network. 
More often than not, real scenario is presented as a multiplex network with common (anchor) users active in multiple social networks.
In the literature, most existing works study either the intra-link prediction in a single network or inter-link prediction among networks (a.k.a. network alignment), and  consider two learning tasks are independent from each other, which is still away from the fact.
On the representation space, the vast majority of existing methods are built upon the traditional Euclidean space, unaware of the inherent geometry of social networks.
The third issue is on the scarce anchor users. Annotating anchor users is laborious and expensive, and thus it is impractical to work with quantities of anchor users.
Herein, in light of the issues above, we propose to study a challenging yet practical problem of \emph{Geometry-aware Collective Link Prediction across Multiplex Network}. To address this problem, we present a novel contrastive model, \textbf{RCoCo}, which collaborates intra- and inter-network behaviors in Riemannian spaces.
In RCoCo, we design a curvature-aware graph attention network ($\kappa-$GAT),  conducting attention mechanism in Riemannian manifold whose curvature is estimated by the Ricci curvatures over the network.
Thereafter, we formulate intra- and inter-contrastive loss in the manifolds, in which we augment graphs by exploring the high-order structure of community  and information transfer on anchor users.
Finally, we conduct extensive experiments with 14 strong baselines on 8 real-world datasets, and show the effectiveness of RCoCo.
}




\maketitle

\section{Introduction}
Recent years has witnessed a surge of online social networks. For instance, Reddit becomes popular for the discussion on the news. The images in Instgram attracts millions of participants. The uses of Twitter and Facebook interact with each other all over the world.
\textbf{Link prediction} is one of the most fundamental learning tasks in online social network. 

In the literature, a series of link prediction solutions have been proposed, and concretely, there are strong methods based on random walk {\cite{kddDeepWalk14,kddnode2vec16}}, subgraph patterns \cite{DGI,TKDEFang23}, graph neural networks (GNNs) \cite{GNNSwwwWuLLN21} and the recent contrastive learning {\cite{tkdeLiu23}}. 
To date, most of previous works typically study link prediction in a single network.
In fact, social networks nowadays are connected to each other, and cannot be isolated any more.
The real-world scenario is that users join in multiple social networks for different purposes, and thus the common users connect those networks as a multiplex network.
In the context of a multiplex network, a single social network is considered as a layer in the multiplex network, and the common users are named as \emph{anchor user} aligning different layers.
Correspondingly, links connecting users in the same layer are referred to as \emph{intra-links}, while the links connecting anchor users are referred to as \emph{inter-links}.
With intra-links predicted, a more comprehensive network topology helps to reveal the structural invariance among different layers (e.g., common friending pattern), and thus facilitates to identify the anchor users. In turn, with inter-links predicted, the information in the counterpart layers is transferred via the anchor users, and typically provides further clues to study the similarity and interconnection among nodes.

\emph{Rather than link prediction in a single network, in this paper, we propose to study \textbf{the collective link prediction}, considering intra/inter-links in the multiplex network as a whole. To the best of our knowledge, we make the first attempt to introduce generic Riemannian manifold to the multiplex network, matching the structure of each network layers.} However, we are facing several significant challenges.

\begin{table}[]
	\centering
	\caption{A motivated example: the geometry of Facebook, Twitter$_A$, Twitter$_B$, Foursquare, DBpedia$_{\text{CH}}$, DBpedia$_{\text{EN}}$, AMiner and DBLP datasets in terms of $\delta-$hyperbolicity and curvature.}
	\label{tab:example}
	\vspace{0.1in}
	\begin{tabular}{c|cccccc}
		\hline
		Dataset    & Node    & Link       & \multicolumn{1}{c}{$\delta -\mathbf{{hyperbolicity} }$ } & \multicolumn{1}{c}{Curvature} \\ \hline
		Facebook   & 422,291 & 3,710,789  & \multicolumn{1}{c}{2}              & \multicolumn{1}{c}{-1.2}          \\
		Twitter$_A$   & 669,198 & 12,749,257 &    1.5                               &       -1.7                        \\
		Twitter$_B$   & 5,167   & 164,660    &                     2              &             -1.6                  \\
		Foursquare & 5,240   & 76,972     &   2                                &                 -1.8              \\
		DBpedia$_{CH}$ & 66,469  & 153,929    &  7.5                                 &      +2.1                         \\
		DBpedia$_{EN}$ & 98,125  & 237.674    &     8                              &              +1.1                 \\
		AMiner     & 26,386  & 273,476    & 3    &        -2.0                       \\
		DBLP       & 24,352  & 316,565    & 2.5                                  &      -2.1                         \\ \hline
	\end{tabular}
\end{table}

\textbf{Challenge 1: Collaboration of Inter- \& Intra-network Behavior.} 
Collective link prediction literally consists of two sub-tasks, intra-link prediction and inter-link prediction. 
As mentioned above, intra-link prediction has been well studied \cite{kddDeepWalk14,kddnode2vec16,journalsChen21}. 
In recent years, inter-link prediction also attracts considerable research attention in both industry and academic \cite{kddChen20,kddMu16,cikmKongZY13}, which is also known as anchor link prediction, user identity linkage or network alignment.
The rub is that the two learning tasks are independently studied.
We argue that intra- and inter-link prediction are related to each other, which is thereby studied as a whole.
Specifically, with intra-links predicted, a more comprehensive network topology helps to reveal the structural invariance among different layers (e.g., common friending pattern), and thus facilitates to identify the anchor users.
In turn, with inter-links predicted, the information in the counterpart layers is transferred via the anchor users,  and typically provides further clues to study the similarity and interconnection among nodes.
In other words, intra-link prediction boosts  inter-link prediction, and vice versa.
However, to the best of our knowledge, only a few studies {\cite{journalsZhanZY19}}  take into account of both intra- and inter-links. Collective link prediction still largely remains open.

\textbf{Challenge 2: Representation Space.}
The vast majority of existing works study intra-link prediction or/and inter-link prediction in the traditional Euclidean space {\cite{kddDeepWalk14,kddnode2vec16,DBLP:conf/kdd/ZhangT16,icdmBayati09}}.
Very recently, Wang et al. \cite{IEEEWang20}, Sun et al. \cite{DBLP:conf/icdm/0008Z0WDSY20,cikm0008YPY22} introduce hyperbolic space \footnote{We use the terminology of space and manifold interchangeably throughout this paper. (We say Euclidean space, and do not use manifold in this case.)} to inter-link prediction, while Bai et al. study intra-link prediction in hyperbolic space \cite{DBLP:conf/www/BaiNZZY23} introduce hyperbolic space.
A fundamental question is that \emph{which space is an appropriate representation space, and more concretely, whether or not the representation spaces of different layers are the same.} 
Theoretically, in Riemannian geometry, there exists three types of isotropic spaces: 1) the negatively curved hyperbolic space, where the surface in the manifold  bends inward (e.g., hyperboloids), 2) the positively curved hyperspherical space, where the surface bends outward, and 3) the flat Euclidean space of zero curvature \cite{petersen2006riemannian}.
Each geometry is associated with a class of structures, e.g., hyperbolic space implies tree-like structures, hyperspherical space for cyclic structures, and Euclidean space for grid structures \cite{icmlcurvatureGCN20}.
Thus, when the predefined  space (either the Euclidean or hyperbolic space) is not in line with the underlying geometry of the network, the expressiveness of the learning model is inevitably weakened.
Empirically, we investigate the geometry of several real-world networks in terms of $\delta-$hyperbolicity \cite{nipsChamiYRL19,imChenFHM13} and curvature, where curvature is the constant curvature, and show the results in Table \ref{tab:example}. \footnote{Twitter$_A$ and Twitter$_B$ are two different subsets of the social network of Twitter.}
According to the Gromov group theory, a smaller value of $\delta-$hyperbolicity  implies the corresponding manifold is more similar to the standard hyperbolic space \cite{DBLP:conf/icdm/0008Z0WDSY20}.
The sign of curvature determines the shape of manifolds, as introduced above.
We summarize the key observations  in Table \ref{tab:example}:  different social networks tend to have different geometry, and thus the representation spaces of different layers are not the same in the multiplex network. Hence, given the complexity of internal relationships within social networks, it becomes challenging to determine directly which type of space, whether Euclidean or hyperbolic, is appropriate for embedding the graph.

\textbf{Challenge 3: Learning with Scarce Anchor Users.}
Most inter-link prediction (network alignment) methods rely on quantities of ``anchor users'' given in prior.
However, annotating anchor users requires label information in both social networks, which is costing and sometimes impossible in practice.
Also, the error in anchor users tends to propagate throughout the network, misleading  network representation as well as alignment.
Recently, graph contrastive learning is introduced to investigate the similarity of the graph itself with some graph augmentations \cite{aaaigrapgcontrast23,wwwGraphContrast21,aaaiContrastive22}.
In the literature, most of contrastive methods work with the traditional Euclidean space and, unlike the image domain, the strategy of graph augmentations still largely remains open.
Euclidean contrastive methods cannot be directly applied to Riemannian manifolds owing to the difference in the operators.
We notice that, it is not until very recently that a few studies \cite{aaaiSunYPWY23,sun2023contrastive,ijcaiSunWYPY23} introduce the Riemannian geometry to graph contrastive learning.
However, none of the existing studies make effort to graph contrastive learning for the multiplex network of different geometries, to the best of our knowledge.

In light of the aforementioned issues, we propose to study a challenging yet practical problem of \emph{Geometry-aware Collective Link Prediction across Multiplex Network}, which considers intra- and inter-link prediction as a whole in multiplex network, where each layer is embedded in the manifold according to the inherent geometry.

To address this problem, we propose a novel Riemannian Contrastive Collective predictor, referred to as \textbf{RCoCo}, to predict intra- and inter-links simultaneously. 
The novelty lies in that RCoCo leverages different manifolds in accordance with network structure of each layer, and is learnt with a intra- and inter- contrastive loss, modeling collective link prediction as a whole.
In RCoCo, we design a curvature-aware graph attention network ($\kappa-$GAT) to learn informative user representation in the manifold.
$\kappa-$GAT adapts to the geometry of each layer via a curvature estimator, which employs the Ricci curvature on the edges to summarize the constant curvature of the network.
With the curvature estimated, $\kappa-$GAT conducts curvature-aware attentional aggregation in the gyrovector space of the manifold. Concretely, we figure out the pairwise attention weight in the tangent space, and aggregate the neighbor's feature with the gyro-midpoint, satisfying the manifold-preserving constraint \footnote{The manifold-preserving constraint requires the output of a Riemannian operator lies in the manifold.}. 
For the Euclidean inputs, we first give the augmented input in the tangent space, and then lift it to the manifold with the exponential map.
Thereafter, we present a intra- and inter- contrastive learning in the manifold.
For intra-contrastive learning, instead of the ad-hoc graph augmentations (e.g., heat diffusion and edge rewriting \cite{icmlHassaniA20}), we explore the community structure in the social network itself for the self-augmentation. Concretely, we contrast the original view to the generated supernode view, where we group the community members as a supernode with I-Louvain algorithm \cite{idaLouvain15}, and compute the similarity between manifold-valued samples.
For inter-contrastive learning, we highlight the anchor users and study the similarity among users of different network layers. 
For each anchor, we maximize the representation agreement between user of one network and its counterpart of the other network in the common tangent space.
In this way, with the few anchors, the knowledge in one network is transferred to the other and Riemannian representation spaces are thereby aligned, modeling intra- and inter-network behaviors collaboratively.
Consequently, intra- and inter-link prediction are mutually enhanced in RCoCo.

Overall, the noteworthy contributions of our work are summarized as follows:
\begin{itemize}
\item \emph{Problem}. 
To the best of our knowledge, this is the first attempt to introduce generic  constant curvature  space (CCS) to the problem of collective link prediction across multiplex network, considering the inherent geometry of each layer of social network.

\item \emph{Methodology}. 
We propose a novel Riemannian contrastive model, RCoCo, in which we first design a $\kappa-$GAT with a curvature estimator for the geometry-aware representation learning. Then, we formulate the intra- and inter-contrastive loss in Riemannian manifolds for learning with scarce annotations.

\item \emph{Experiment}.  
We conduct extensive experiments with 14 strong baselines on 8 real-world datasets, and empirical results testify the  superiority of RCoCo.
Moreover, we conduct the ablation study and discuss the parameter sensitivity to evaluate RCoCo.
\end{itemize}

\textbf{Roadmap.}
The content in the rest of our paper is sketched as follows.
In Section 2, we introduce the preliminaries on Riemannian geometry, and we give formal definition of the studied problem, \emph{Self-supervised Collective Link Prediction across Multiplex Network}. 
We present our novel solution, \textbf{RCoCo}, in Section 3.
To evaluate the proposed model, Section 4 describes our experimental setup,
and empirical results are shown in Section 5.
In Section 6, we elaborate on the related work on intra- and inter-link prediction, graph contrastive learning and Riemannian machine learning on graphs. 
Finally, we summarize our work and highlight key contributions in Section 7.


\section{Preliminaries}

In this section, we first formally review the basic concepts of Riemannian geometry, including Riemannian manifolds, geodesics, and curvature. (The formulas of the operations in our model can be found in  Table \ref{tab:t1}.) Then, we introduce the definition of anchor user and multiplex network to formulate the studied problem. 

\subsection{Riemannian Geometry}
\subsubsection{Riemannian Manifold}
\noindent\textbf{Tangent space.}
Tangent space describes the local property of the Riemannian manifold. At each point $x$ on the manifold $\mathcal{M}$, a tangent space is defined as $T_{x} \mathcal{M}$. Its dimension is the same as that of manifold.
  
\noindent\textbf{Riemannian metrics.} 
An inner product $ \left \langle \cdot , \cdot \right \rangle   _{T_{x}	\mathcal{M}  } : T_{x}\mathcal{M} \times T_{x}\mathcal{M}\to \mathbb{R} $ is defined on the tangent space, and the collection of the product is expressed as Riemannian metric $g$. Endowed with the Riemannian metric, a $n-$dimensional smooth manifold is  said to be a Riemannian  manifold $\left ( \mathcal{M}, g\right )$. 

\noindent\textbf{Geodesic.}
A geodesic is the shortest curve between two points in manifold. In Euclidean space, the geodesic is a straight line. In a Riemannian manifold, we first defined the infinitesimal as: $\mathrm{d}s^{2} = g_{ij} \mathrm{d}x^{i}\mathrm{d}x^{j}$. Then for $\gamma :\left ( \alpha ,\beta  \right ) \longrightarrow \mathcal{M}$, we define the geodesic on $\left ( \mathcal{M}, g\right )$: $ L\left ( \gamma  \right ) = arg_{\gamma} min \int_{\alpha }^{\beta } \left \| \gamma '{\left ( t \right ) }  \right \|_{g}dt  $. 

\noindent\textbf{Riemannian vector modulus.}
We first give the formulation of the inner product with the metric: $g\left ( X,Y \right ) = g_{ij} X^{i} Y^{j} $, then the modulus length and angle of the tangent vector are also defined accordingly: $\left \| X \right \| =g\left ( X,X \right ) ^{\frac{1}{2} } $, $\cos \angle \left ( X,Y \right ) =\frac{g\left ( X,Y \right ) }{\left \| X \right \|\cdot  \left \| Y \right \| } $, where $\left \| X \right \|$ is a norm in Euclidean space.

\noindent\textbf{Exponential and logarithmic mapping.}
For the Riemannian manifold $(\mathcal{M}, g)$ and the tangent space, exponential mapping maps the vector $\textbf{v}$ in tangent space back to manifold, defined as $exp_{x}^{\kappa}(\textbf{v}): T_{x} \mathcal{M}\longrightarrow  \mathcal{M}$. Logarithmic mapping maps the vector in manifold back to tangent space, defined as $log_{x}^{\kappa}(\textbf{v}): \mathcal{M} \longrightarrow  T_{x} \mathcal{M}$. Logarithmic mapping can be regarded as the inverse operator of exponential mapping, and vice versa.

\noindent\textbf{Parallel transport.}
For two points on a manifold, parallel transport is a vector of a point in tangent space along a geodesic line to transport another point. Define parallel transport from point $x$ to $y$ as $PT_{x\to  y}  : T_{x} \mathcal{M}\longrightarrow T_{y} \mathcal{M}$. 

\noindent\textbf{Gyrovector space.}
Gyrovector space is an elegant mathematical formalism defined on an open ball. 
Riemannian manifold is not a typical vector space, where Euclidean vector operations cannot be applied, e.g., vector addition is not closed in the manifold,  $\mathbf x + \mathbf  y \notin \mathcal M$, for $\mathbf x, \mathbf  y \in \mathcal M$. 
Gyrovector space provides Euclidean-like and non-associative vector operations. As for the non-associative property, we have $\mathbf x \oplus \mathbf y \neq \mathbf y \oplus \mathbf x$, for $\mathbf x, \mathbf  y$ in the manifold. 
($+$ and $\oplus$ denote vector addition in Euclidean and Riemannian manifold, respectively.)

\subsubsection{Curvature}

When a vector $v$ moves in the $x^\mu$ direction and then $x^\beta$ is different from the vector moving along $x^\beta$ and then $x^\mu$, then the space has curvature. 
Curvature can be divided into the following three categories according to positive and negative:

\textbf{Positive curvature}: In the case of positive curvature, the curve on the Riemannian manifold tends to bend outward, just like a sphere. Examples of positive curvature include spherical and ellipsoidal surfaces.
\textbf{Negative curvature}: In the case of negative curvature, the curve on the Riemannian manifold tends to bend inward, just like hyperboloids. Examples of negative curvature include hyperboloids.
\textbf{Zero curvature}: In the case of zero curvature, the curves on the Riemannian manifold do not bend, similar to a straight Euclidean space.

Curvature is an important concept to describe manifolds, the Riemannian curvature tensor formula is given below:
\begin{equation}
R_{i,j,k}^{l}=\partial _{k} \Gamma_{ij}^{l}-\partial _{j} \Gamma_{ik}^{l}+\Gamma_{ij}^{p} \Gamma_{pk}^{l}-\Gamma_{ik}^{p} \Gamma_{pj}^{l}
\end{equation}
$\Gamma_{ij}^{l}$ is the component of the Riemannian contact, and $\partial _{k}$ denotes finding the partial derivative. This formula is obtained by combining the first partial derivatives and squares of the contact.

\subsection{Problem Statement}

A social network is described as a tuple of $G=(\mathcal V, \mathcal E)$, where $\mathcal{V}=\{ v_{i} \}$  and $\mathcal{E}=\{ (v_{i}, v_{j}) \} \subset \mathcal{V} \times \mathcal{V}$ denote the user account set and edge set, respectively. Each account  is coupled with a $d-$dimensional feature vector $\mathbf x \in \mathbb R^d$. 

\newtheorem*{def1}{Definition (Multiplex Network)} 
\begin{def1}
A multiplex network $\mathcal{G}=(G_1, G_2, \cdots, G_K)$ is constructed with a set of aligned network, and a network $G_k$ is referred to as the $k-$th layer of the multiplex network.
\end{def1}
Typically, multiplex network layers are fully aligned. However, we relax the full alignment in typical multiplex networks, and consider partially aligned layers in the real-world scenario. In our context, each social platform serves as a layer in the multiplex network.
In this paper, we consider a two-layer multiplex network, denoted as $G_{s}$ and $G_{t}$, and note that our solution can be easil extended to the multiple-layer cases. There exists two types of links in the multiplex network: intra-links and inter-links.

\newtheorem*{def3}{Definition (Intra- \& Inter-Link)} 
\begin{def3}
For a multiplex network $\mathcal{G}=(G_1, G_2, \cdots, G_K)$ with $G_k=(\mathcal V_k, \mathcal E_k)$, intra-links are given in edge sets $\mathcal V_k$, connecting the nodes in the same network layer. Inter-links $\mathcal{D}_{st} \subset \mathcal{V}_{s} \times \mathcal{V}_{t}$ connect the nodes in different network layers $s \neq t$, describing the alignment of the multiplex network.
\end{def3}
\noindent 
Alternatively, intra-links in network $G_k$ are denoted by the adjacency matrix $\textbf{A}_k$. $\textbf{A}_k$[$i^{th}$, $j^{th}$] $=$ $1$ if there is a link between nodes $v_i$ and $v_j$.
In this paper, we only consider inter-links within the same user identity. We define a matrix $\textbf{D}$ of size $N_s$ by $N_t$, where $N_s$ and $N_t$ denote the number of nodes in $G_s$ and $G_t$ respectively. The matrix $\textbf{D}$ records the alignment between networks $G_s$ and $G_t$.

\newtheorem*{def2}{Definition (Anchor User)} 
\begin{def2}
For any $v_i \in \mathcal V^s$ and $v_j \in \mathcal V^t$, $v_i $ and $v_j$ are said to be the anchor user if and only if $\phi(v_i)=\phi(v_j)$, both account point to the same user identity in the real world. Correspondingly, the inter-link connects the anchor user is also referred to as anchor link.
\end{def2}
\noindent We consider that a user owns \emph{at most one account} in a social platform, and we perform account fusion if there exists multiple accounts of the same user. In the rest of this paper, we interchangeably use the term of anchor link and inter-link.

With the concepts of multiplex network and intra- \& inter-links, we formulate the studied problems as follows.
\newtheorem*{def4}{Problem  Definition (Geometry-aware Collective Link Prediction across Multiplex Network)} 
\begin{def4}
Given a multiplex network $\mathcal{G}=(G_s, G_t)$ with $G_s=(\mathcal V_s, \mathcal E_s)$ and $G_t=(\mathcal V_t, \mathcal E_t)$, the aim of this problem is to find a neural predictor which is able to collectively predict two types of links
\begin{itemize}
\item intra-links, the high-probability future connection between nodes in the same network layer, and 
\item inter-links, the alignment between the two layers with the scarce anchor annotations.
\end{itemize}
The neural predictor operates in the generic manifold, and is aware of the underlying geometry of the network structure in each layer.
\end{def4}
\noindent Different from the previous works, we are the first to highlight layer-wise geometry for the collective link prediction across multiplex network. That is, each layer works with its own geometry in accordance with the network structure.

\begin{table}
    \caption{Important Notations.}
\vspace{0.1in}
    	\label{tab:t2}
    	\centering
    	\begin{tabular}{c | l}
    		\hline
    		Symbol & Description \\
    		\hline
    		$\mathbb H, \mathcal{M}$ & Hyperbolic, Riemannian manifold\\
    		$ \mathcal{M}^d_\kappa$ & $\kappa-$stereographical model of Riemannian manifold\\
    		$\kappa, d$ & Constant curvature and dimension, respectively \\
    		$\lambda^\kappa_{m}$ & Conformal factor of the point $ m$  in  manifold $\mathcal{M}^d_\kappa$\\
    		$Ricci^{\alpha}( x,  y)$  & Ricci curvature between points $ x$ and $ y$\\
    		$T_{\boldsymbol x}\mathcal M$ & Tangent space of $\boldsymbol x$ \\
    		$G$ & Social network\\
    		$\textbf{W} $ & The weigth matrix of $G$\\
    		$\textbf{X} $ & The characterstic matrix of $G$\\
    		$\textbf{A} $ & The adjacency matrix of $G$\\
    		$\textbf{D}$ & The Anchor links matrix between network $G$$_s$ and $G$$_{t}$ \\
    		$\mu$ & Gyromidpoint\\
    		$\eta$ & Overlap ratio\\
    		\hline
    	\end{tabular}
    	\label{notation}
\end{table}

\section{Our Approach: 	RCoCo}

\begin{figure*}[htbp]
	\centering
	\includegraphics[width=13cm,height=5cm]{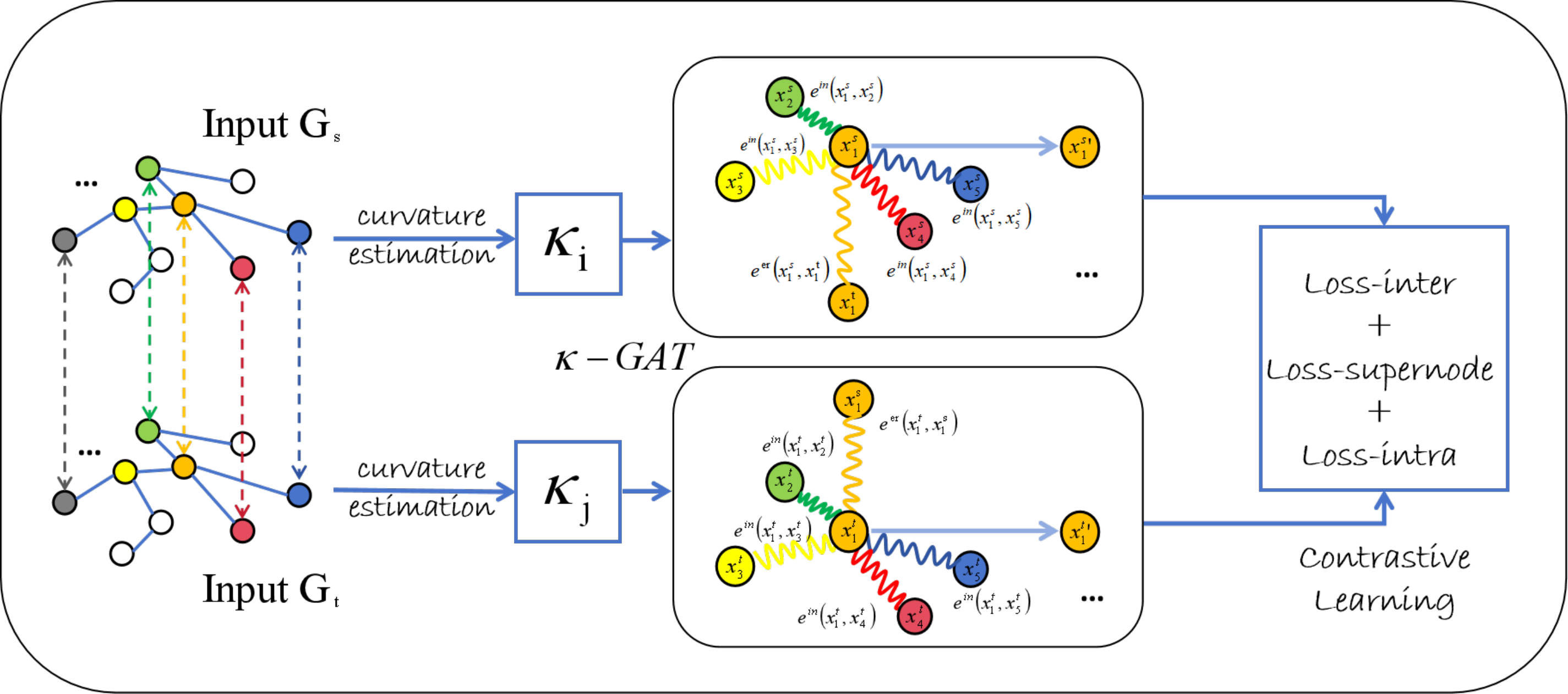}
	\caption{Overall architecture of RCoCo. For two networks with anchor users (the dotted line in the figure represents the anchor user), we first designed a curvature estimator to adapt the geometry of each graph which employs the Ricci curvature. Then with the curvature estimated $\kappa_{i}$, $\kappa$-GAT conducts intra- and inter- network attention aggregation in the manifold, where $e^{in}(x^{s},x^{s})$ in figure represents intra network attention, and $e^{er}(x^{s},x^{t})$ represents intra network attention. Finally, we conduct intra- and inter-contrastive learning in the manifold via three loss functions.}
	\label{figure}
\end{figure*}

To address the problem above, we present a novel Riemannian Contrastive Collective predictor (\textbf{RCoCo}). In brief, RCoCo updates the node representation with neighbor nodes through $\kappa$-GAT. With the curvature estimated, $\kappa$-GAT conducts curvature-aware attentional aggregation in the gyrovector space of the manifold. We illustrate the architecture of RCoCo in Fig. \ref{figure}. To resolve the issue of  scarce anchor annotations, RCoCo presents a intra- and inter-contrastive learning in the manifold. 

Next, we will elaborate on the proposed component of RCoCo.

\subsection{Graph Attention in Euclidean Space}
We start with introducing the Graph Attention Network (\textbf{GAT} \cite{iclrGAT18}) in Euclidean space.  It updates the nodes representation by aggregating the characteristics of its neighboring nodes. 
The attention mechanism is defined as follows: $Attention =\Sigma similarity \ast Value_{i}$. The two main key steps of GAT are: 1) to calculate the attention coefficient through the attention mechanism, 2) to aggregate the node vectors via the calculated attention coefficient. 
The focus of GAT is to calculate the graph attention correlation coefficient, Velickovic et al. \cite{iclrGAT18} provide a method to calculate the graph attention coefficient, as follows.
\begin{equation}
	e_{ij} = LeakyReLu\left ( w^{T}\left [ \textbf{W}h_{i } \left |  \right | \textbf{W}h_{j }\right ]   \right ) 
\end{equation}

Euclidean GAT  has the following drawbacks. 1) When the social network does not present dominant Euclidean structure, Euclidean GAT tends to present large embedding distortion. 2) More importantly,  the attention mechanism a single graph cannot model the complex pattern in multiple network for collective link prediction. 

\subsection{Graph Attention in Riemannian Manifold}
To bridge the gaps above, in RCoCo, we design a curvature-aware graph attention network ({$\kappa$}-GAT) to learn informative user representation in the manifold.

\noindent\textbf{$\kappa$-stereographic model.}
Before discussing the representation and aggregation of graph nodes, we give a manifold model and introduce the related operations. In this paper, we opt $\kappa$-stereographic model, as it is able to describe spaces with either positive or negative curvature.
Specifically, $\kappa$-stereographic model is a smooth manifold defined as $\mathcal{M}_{k}^{d}$, for curvature $\kappa \in \mathbb{R}$, the model is described below:
\begin{equation}
	\mathcal{M}_{k}^{d} =\left \{ m\in \mathbb{R} ^{d} \mid  -\kappa  \left \| m \right \| _{2}^{2}  < 1\right \} , 
\end{equation}
where $d\ge 2$. The Riemannian metric is given as $g_{m}^{\kappa } = \left ( \lambda _{m}^{\kappa }   \right ) ^{2}\pmb{I}$, where the conformal factor $\lambda _{m}^{\kappa }$ is defined as: 
$	\lambda _{m}^{\kappa } =\frac{2}{ 1+\kappa \left \| m \right \|^{2}  }$.

In detail, the model is defined on the unrestricted domain of  $( \mathbb{R} ^{d}) $ when $ \kappa \ge 0$.  The model is the open ball with a radius of $ \frac{1}{\sqrt{- \kappa } } $ when $  \kappa <  0$.

\begin{table}[t]
	\centering
	\caption{Summary of the operations with unified formalism.}
	\label{tab:t1}
\vspace{0.1in}
{
		\begin{tabular}{|c|c|}
			\hline
			\textbf{Operation}  & \textbf{Unified gyrovector formalism in $\mathcal{M}_{\kappa}^{d}$}   \\
			\hline
			Distance Metric 
			& $d_{\mathcal{M}}^{\kappa}\left(\mathbf{x}, \mathbf{y}\right) = \frac{2}{\sqrt{|\kappa|}}\tan_{\kappa}^{-1}\left(\sqrt{|\kappa|} \left\|-\mathbf{x} \oplus_{k} \mathbf{y}\right\|_{2}\right)$
			  \\
			\hline
			Gyrovector Addition 
			& $\mathbf{x} \oplus_{\kappa} \mathbf{y} =\frac{\left(1-2 \kappa\langle\mathbf{x}, \mathbf{y}\rangle-\kappa\|\mathbf{y}\|_{2}^{2}\right) \mathbf{x}+\left(1+\kappa\|\mathbf{x}\|_{2}^{2}\right) \mathbf{y}}{1-2 \kappa\langle\mathbf{x}, \mathbf{y}\rangle+\kappa^{2}\|\mathbf{x}\|_{2}^{2}\|\mathbf{y}\|_{2}^{2}}$ 
			  \\
			Gyrovector Scaling & $r \otimes_{\kappa} \mathbf{x} =\frac{1}{\sqrt{\kappa}} \tanh \left(\kappa \tanh ^{-1}(\sqrt{\kappa}\|\mathbf{x}\|_{2})\right) \frac{\mathbf{x}}{\|\mathbf{x}\|_{2}}$ 
			  \\
			Matrix-Vector Multiplication & $\boldsymbol{M} \otimes_{\kappa} \mathbf{x} =(1 / \sqrt{\kappa}) \tanh \left(\frac{\|\boldsymbol{M} \mathbf{x}\|_2}{\|\mathbf{x}\|_2} \tanh ^{-1}(\sqrt{\kappa}\|\mathbf{x}\|_2)\right) \frac{\boldsymbol{M} \mathbf{x}}{\|\boldsymbol{M} \mathbf{x}\|_2}$
			 \\
			$\kappa$-Right-Multiplication &  $\boldsymbol{X} \otimes_{\kappa} \boldsymbol{W} =\exp_{\boldsymbol{\nu}}^{\kappa}(\log _{\boldsymbol{\nu}}^{\kappa}(\boldsymbol{X})\boldsymbol{W}) $
			\\
			\hline
			Exponential Map & $\exp _{\mathbf{x}}^{\kappa}(\mathbf{v}) =\mathbf{x} \oplus_{\kappa}\left(\tan _{\kappa}\left(\sqrt{|\kappa|} \frac{\lambda_{\mathbf{x}}^{\kappa}\|\mathbf{v}\|_{2}}{2}\right) \frac{\mathbf{v}}{\|\mathbf{v}\|_{2}}\right)$ 
			 \\
			Logarithmic Map & $\log _{\mathbf{x}}^{\kappa}(\mathbf{y}) =\frac{2}{\lambda_{\mathbf{x}}^{\kappa}\sqrt{|\kappa|}} \tan _{\kappa}^{-1}\left\|-\mathbf{x} \oplus_{\kappa} \mathbf{y}\right\|_{2} \frac{-\mathbf{x} \oplus_{\kappa} \mathbf{y}}{\left\|-\mathbf{x} \oplus_{k} \mathbf{y}\right\|_{2}}$ 
			    \\
			\hline
			& $\tan_{\kappa}\left(\mathbf{x}\right) = \begin{cases}
				\tanh\left( \mathbf{x}\right), & \kappa < 0, \\
				\tan\left( \mathbf{x}\right), & \kappa > 0.
			\end{cases}$ 
			 \\
			Curvature Trigonometry& $\cos_{\kappa}\left(\mathbf{x}\right) = \begin{cases}
				\cosh \left( \mathbf{x}\right), & \kappa < 0, \\
				\cos \left( \mathbf{x}\right), & \kappa > 0.
			\end{cases}$ 
			  \\
			& $\sin_{\kappa}\left(\mathbf{x}\right) = \begin{cases}
				\sinh \left( \mathbf{x}\right), & \kappa < 0, \\
				\sin\left( \mathbf{x}\right), & \kappa > 0.
			\end{cases}$
			  \\
			\hline
	\end{tabular}}

\end{table}

\noindent\textbf{Input Layer.}
We introduce an input layer to transform Euclidean inputs to the vectors in the manifold.
First, we give the augmented input in the tangent space. 
Specifically, we refer to $\left ( \sqrt{\kappa },0,0,\cdots , 0\right )$  as the origin of the manifold, denoted as $\boldsymbol \nu $. 
For vector $\boldsymbol{x}\in \mathbb{R} ^{d} $, we have its corresponding vector $\boldsymbol{v} ^{ \boldsymbol \nu   } $ in the tangent space $T_{ \boldsymbol \nu   }  \mathcal{M} ^{d,\kappa } $, and $\left \langle \boldsymbol{v} ^{ \boldsymbol \nu   } ,\boldsymbol \nu   \right \rangle =0$. 
The augmented input is  constructed as $\boldsymbol{v}  =(0,\boldsymbol{x})$. 
Then, we lift it to the manifold with the exponential map at $\boldsymbol{v} ^{\boldsymbol  \nu  } $.  The transformed vector in Riemannian space is given as
\begin{equation}
	{\boldsymbol \nu} _{\mathbb{R}\to \mathcal{M}  }^{\kappa } \left ( x\right )  = exp_{\boldsymbol \nu  }^{\kappa }\left ( \boldsymbol{v} ^{\boldsymbol \nu  }  \right ),
\end{equation}

\noindent\textbf{Feature Transformation.}
The feature transformation in Riemannian space is realized by left-multiplication, defined as $\otimes ^{\kappa } $. Now, we can transform a d-dimensional vector to a g-dimensional vector via matrix multiplication in Riemannian space, which is defined as follows:
\begin{equation}
	\boldsymbol{W}\otimes ^{\kappa } x=exp_{\boldsymbol{\nu}  }^{\kappa } \left ( \boldsymbol{W}log_{\boldsymbol{\nu} }^{\kappa }\left ( x \right )  \right ) 
\end{equation}
where $\boldsymbol{W}$ is the d$\times $g-dimensional weight matrix.

\noindent\textbf{Riemannian Attention Layer.}
For a node needs to be aligned, we first go through the intra-network attention mechanism. 
We update the representations of the node by aggregating the neighbor's feature. Since the importance of neighbors usually different, we introduce the intra-network attention: $e^{in}\left ( x_{i}, x_{j}  \right )$, to represent the magnitude of the influence of node $x_{j}$ on node $x_{i}$ in the network. We first lift nodes in tangent spaces via logarithmic mapping, and the attention parameters $e^{in}\left ( x_{i}, x_{j}  \right )$ in Riemannian space are modeled as follows:
\begin{equation}
	e^{in}\left ( x_{i}, x_{j}  \right ) =\sigma \left ( \beta _{in} \left ( \boldsymbol{W}^{x}log_{\boldsymbol{\nu}  }^{\kappa } \left ( x_{i}  \right ) \left |  \right | \boldsymbol{W}^{x}log_{\boldsymbol{\nu}  }^{\kappa } \left ( x_{j}  \right ) \right )  \right ) ,
\end{equation}
where $\boldsymbol{W}^{x}$ is the weight matrix, which is used for the feature transformation of nodes. $\sigma$ is the sigmoid activation function. $\beta _{in}$ is the weight vector, which models the importance parameterized. 

In addition to intra-network attention, as information across the network also plays a important role, we introduce the inter-network graph attention. 
We define anchor links as $\mathcal{C} \left ( x_{i}  \right ) =\left \{ y_{j } \mid \left ( x_{i} ,y_{j}  \right ) \in \textbf{D}^{x,y}   \right \} $, where $\textbf{D}^{x,y}$ is the anchor links matrix in Graph ${G}_{x}$ and ${G}_{y}$. For $x_{i}$ in network ${G}_{x}$, its inter-network attention parameter is as follows, which represents the influence of the corresponding node on $x_{i}$.
\begin{equation}
	e^{er} \left ( x_{i} ,y_{j}  \right ) = \sigma \left ( \beta _{er}  \left ( \boldsymbol{W}^{x} log_{\boldsymbol{\nu} }^{\kappa } \left ( x_{i}  \right )  \right ) \left |  \boldsymbol{W}^{xy} log_{\boldsymbol{\nu} }^{\kappa } \left ( y_{j}  \right )\right | \right ) 
\end{equation}
We use softmax to normalize the attention mechanism parameter $e^{in}$, $e^{er}$, and obtain the attention weight $E$ as follows, where $\mathcal{N}$ is the set of neighbor nodes:
\begin{equation}
	E^{in}_{x_i,x_j} =\frac{exp\left ( e^{in}\left ( x_{i}, x_{j}  \right ) \right ) }{\Sigma _{x_{k}\in \mathcal{\mathcal{N} }\left ( x _{i}\right )  } exp\left ( e^{in}\left ( x_{i}, x_{k}  \right ) \right )+\Sigma _{y_{k}\in \mathcal{C}\left ( x _{i}\right )  } exp\left ( e^{er}\left ( x_{i}, y_{k}  \right ) \right ) }
\end{equation}
\begin{equation}
	E^{er}_{x_i,y_j} =\frac{exp\left ( e^{er}\left ( x_{i}, y_{j}  \right ) \right ) }{\Sigma _{x_{k}\in \mathcal{\mathcal{N} }\left ( x _{i}\right )  } exp\left ( e^{in}\left ( x_{i}, x_{k}  \right ) \right )+\Sigma _{y_{k}\in \mathcal{C}\left ( x _{i}\right )  } exp\left ( e^{er}\left ( x_{i}, y_{k}  \right ) \right ) }
\end{equation}

\subsection{Curvature Estimation}
{$\kappa$-}GAT adapts to the geometry of each layer via a curvature estimator, which employs the Ricci curvature on the edges to summarize the constant curvature of the network.

Ricci Curvature is geometrically used as a function of curvature to control the rate at the volume of the ball grows.
Ollivier et al. \cite{2009Ricci} defines Ricci curvature. When given the mass distribution function $m_{u}^{\alpha} \left ( v \right )$ of the node $u$ on a graph,
\begin{equation}
	m_{u}^{\alpha} \left ( v \right ) =
	\left\{
	\begin{aligned}
		& \alpha , && v= u, \\
		& \left ( 1-\alpha  \right ) \frac{1}{\left|\mathcal{N}\left(u\right)\right|}, && v\in \mathcal{N}\left(u\right), \\
		& 0, && \text{otherwise},
	\end{aligned}
	\right.
\end{equation}
where $\mathcal{N}$ is the set of neighbor nodes. Then the Ricci curvature of a pair of nodes $(i,j)$ is as follows:
\begin{equation}
Ricci^{\alpha } \left ( i,j \right ) =1-\frac{W\left ( m_{i}^{\alpha } , m_{j}^{\alpha } \right ) }{ d\left ( i,j \right ) } ,
\end{equation}
$ W\left ( m_{i}^{\alpha } , m_{j}^{\alpha } \right )$ is the Wasserstein distance between $i$ and $j$, and $d\left ( i,j \right )$ is the graph distance. 
Incidentally, Wasserstein distance is a measure of the probability distributions difference. It has the ability to transform one distribution into another while preserving its geometric characteristics.

Correspondingly, the Ricci curvature of a node is derived as $Ricci^\alpha(i)=\frac{1}{degree_i}\sum_{j}Ricci^\alpha(i,j )$, where $degree_i$ represents the degree of point $i$. In practice, we randomly sample nodes from a uniform distribution, and compute the expectation of node curvatures, yielding the network curvature.


\begin{algorithm}
		\SetKwData{Left}{left}\SetKwData{This}{this}\SetKwData{Up}{up}
		\SetKwFunction{Union}{Union}\SetKwFunction{FindCompress}{FindCompress}
		\SetKwInOut{Input}{input}\SetKwInOut{Output}{output}
		
		\Input{Source social network ${G}_{s}$ and its adjacency matrix $\textbf{A}^{s}$, Target social network ${G}_{t}$ and its adjacency matrix $\textbf{A}^{t}$, Anchor links the set of users $\mathcal{P}$}
		\Output{Nodes represent matrices $\textbf{X}^{s}$ and $\textbf{X}^{t}$,  ${G}_{s}$ and ${G}_{t}$'s user alignment list collections}
		\BlankLine
		//  Initialize
		
		\For{network ${G} \in \left (  {G}_{s} , {G}_{t}\right ) $}{
			Initialize ${G}_ {2} \gets {G};$
			
			\For{$l \gets 1,2$}{Initalize $\textbf{W}_{0}^{l}$ and $\textbf{W}_{1}^{l}$;}}
			 
			\For{alternation $\gets$ 1, \dots, $N^{alt}$}{ ${G}_{s,2} \gets {G}_{s}$;
				
				${G}_{t,2} \gets {G}_{t}$;
				 
			\For{network ${G}_{2} \in \left (  {G}_{s,2} , {G}_{t,2}\right ) $}{${G} \gets I-Louvain({G}_{2})$}
			\For{epoch $\gets$ 1, \dots, $N^{epo}$}{
				\For{$l \gets 1, 2$}{\For{network ${G} \in \left (  {G}_{s} , {G}_{t}\right ) $}{
				//  Intra-network Contrastive Learning
				
				Generate A and X;
						
				Compute $\mathcal{L}_{in}$ and $\mathcal{L}_{n2s}$;
				
				Update $\textbf{W}_{0}^{l}$ and $\textbf{W}_{1}^{l}$ using $\mathcal{L}_{in}$ and $\mathcal{L}_{n2s}$;
				
				//  Inter-network Contrastive Learning
				
				Generate A and X;
				
				Compute $\mathcal{L}_{er}$;
				
				Update $\textbf{W}_{0}^{l}$ and $\textbf{W}_{1}^{l}$ using $\mathcal{L}_{er}$;
			}

	}}
			
			}

		\caption{RCoCo Method}\label{algo_disjdecomp}
\end{algorithm}

\subsection{Contrastive Intra- \& Inter-link Prediction}
Thereafter, we present a intra- and inter-contrastive learning in the manifold. Specifically, for intra-contrastive we use I-Louvain to group the community members as a supernode, and contrast the node view and supernode view. For inter-contrastive, we maxmize the representation agreement between anchor users. We summarize the learning process of RCoCo with contrastive loss in Algorithm \ref{algo_disjdecomp}.   

\noindent\textbf{Contrastive Learning.}
The basic idea of graph Contrastive Learning is to compare two or more datasets, find out the differences and commonalities, and learn the connections between them.

Contrastive Learning methods usually have three main steps: 1) the first step is to augment the graph data to get new views. 2) Then encode view to obtain the representation vectors of nodes and graphs. The same sample in both views is a positive example, and different samples are negative. 3) Finally, calculate the similarity between samples via the similarity function \cite{icmlChenK0H20}, and positive examples exhibit high similarity, while negative examples demonstrate the opposite.
The contrastive loss of a positive pair $(x_{i},x_{j})$ is defined as:
\begin{equation}
	l\left ( x_{i},x_{j}\right )  = - \log_{}{\frac{exp\left ( \varphi \left ( \boldsymbol{x}_{i} ,\boldsymbol{x}_{j} \right ) \right ) }{ \sum_{q=1}^{2N}exp\left ( \varphi \left ( \boldsymbol{x}_{i} ,\boldsymbol{x}_{q} \right ) \right )  }} 
\end{equation}


\noindent\textbf{Graph Augmentation \& Pseudo-Labels.} 
I-Louvain \cite{idaLouvain15} is an attribute graph clustering algorithm. We group the community members as a supernode with I-Louvain algorithm and determine the most appropriate label.
For network $G$, we use I-Louvain to maximize module $\mathcal{Q} $, defined as $\mathcal{Q}= \mathcal{Q}_{iner} +  \mathcal{Q}_{NG} $, 
\begin{equation}
	\mathcal{Q}_{NG}  = \frac{1}{2m}\sum_{i,j}  \left ( \left [ \boldsymbol{A} \right ]_{ij} -\frac{d_{i} d_{j}}{2m} \right )\cdot   \delta \left ( v_{i} ,v_{j}  \right ) , 
\end{equation} 
\begin{equation}
	\mathcal{Q}_{iner}=\sum_{i,j}\left ( \frac{iner\left (  \boldsymbol{X},x_{i}  \right ) \cdot iner\left ( \boldsymbol{X},x_{j}  \right )}{\left ( 2N\cdot iner\left ( \boldsymbol{X} \right )   \right ) ^{2}} -\frac{d_{\kappa } \left ( x_{i}-x_{j}  \right ) ^{2} }{2N\cdot iner\left ( \boldsymbol{X} \right ) }  \right )\cdot \delta \left ( v_{i} ,v_{j}  \right )    
\end{equation} 
where $\boldsymbol{X}$ is the characteristic matrix of the network $G$, $\textbf{A}$ is the adjacency matrix of $G$, $d_{i}$ is the degree of $v_{i}$, and $m$ is the number of edges. $\delta \left ( v_{i} ,v_{j}  \right )$ is a function that indicates whether $v_{i}$ and $v_{j}$ are divided into the same block. $\mu$ is the gyromidpoint of $\boldsymbol{X}$, which is \begin{equation}
	\mu\left (\boldsymbol{X}_{N} ;E \right ) = \frac{1}{2} \oplus \left ( \sum_{l=1}^{N}  \frac{E _{il}\lambda _{{\boldsymbol{X}_{i}}}^{\kappa }  }{ {\textstyle \sum_{j=1}^{N}E _{ij}\left ( \lambda _{{\boldsymbol{X}_{i}}}^{\kappa }  -1 \right ) } }\boldsymbol{X}_{i}  \right ) .
\end{equation} 
$iner\left ( x \right ) =  {\textstyle \sum_{i}^{N}} d_{\kappa } \left ( x_{i} -\mu  \right )^{2}  $ is the sum of the differences between all points and midpoint $\mu$, representing the inertia of $G$. $iner\left ( \boldsymbol{X},x_{i} \right ) =  {\textstyle \sum_{j}^{N}} d_{\kappa } \left ( x_{i} - x_{j}  \right )^{2}  $ is the sum of the difference between all points and $x_{i}$, representing the inertia through $x_{i}$.



\noindent\textbf{Intra-network Contrastive Learning.}
\begin{figure*}[htbp]
	\centering
	\includegraphics[width=13cm,height=4cm]{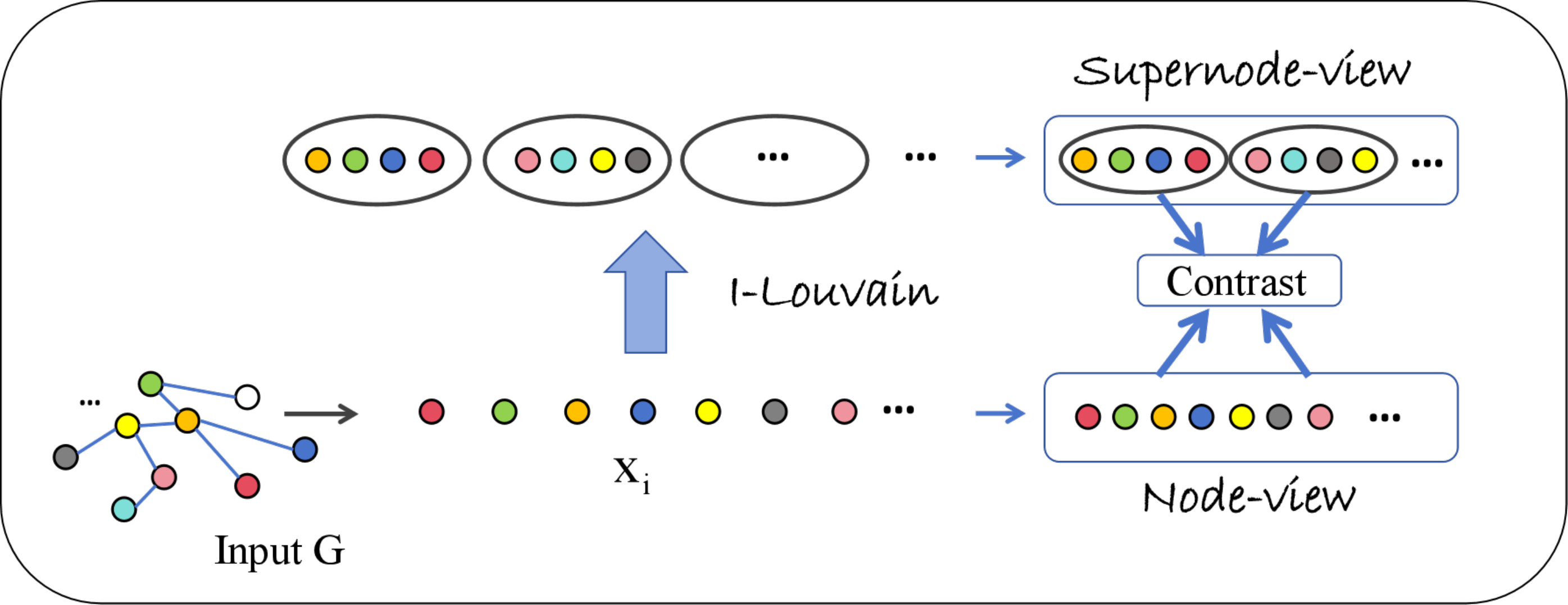}
	\caption{Node-Supernode contrastive Learning. We first group the community members as a supernode with I-Louvain algorithm, and then contrast the original node view and generated supernode view.}
	\label{Fig.2}
\end{figure*}

Our model identifies nodes in the network by maximizing the similarity gap between pairs of nodes, that is, maximizing the similarity of positive examples and minimizing the similarity of negative examples. 

For intra-contrastive learning, instead of the adhoc graph augmentations, we explore the community structure in the social network itself for the self-augmentation. Concretely, For each network $G$, we generate its supernode view $G$${}' $ via I-Louvain algorithm, and then set up two levels of comparison: 1) Node-Node level performs the self-contrast in $G$. 2) Node-Supernode level contrasts the node view $G$ and the supernode view  $G$${}' $. As shown in Fig. \ref{Fig.2}, we contrast the original view to the generated supernode view, and compute the similarity between manifold-valued samples. Thus, the intra-network loss is the sum of the two loss functions 1) and 2).

	\textbf{Node-Node: }In the self-contrast, node $x_{i}$ is a negative sample of  node $x_{j}$ if $i$ is not equal to $j$.
The contrastive loss of a node pair  is defined as follows:
\begin{equation}
	l_{in} \left ( v_{i} , v_{j}\right )  = - \log_{}{\frac{exp\left ( sim_{\kappa}\left ( x_{i} ,x_{j} \right )  \right ) }{\sum_{g=1}^{N}exp\left ( sim_{\kappa}\left ( x_{i} ,x_{g} \right ) \right ) }} 
\end{equation}
where $sim_{\kappa}(x,y)=\left \langle log_{\nu }^{\kappa  } (x),log_{\nu }^{\kappa } (y) \right \rangle  $. The contrastive loss of node-node level is given as follows: 
\begin{equation}
	\mathcal{L}_{in}   = \frac{1}{2N} \sum_{k=1}^{N} \frac{1}{\left | \mathcal{X}\left ( k \right )   \right | }\sum_{\left ( i,j \right )\in \mathcal{X} \left ( k \right )  }^{}\left ( l_{in}  \left ( v_{i},v_{j}  \right ) +  l _{in} \left ( v_{j},v_{i}  \right )\right ),   
\end{equation}
where $N$ is the number of nodes in the network.

For network $G$, we use I-Louvain to generate supernode view $G$${}' $. For each node in $G$, 1) we calculate the $\Delta \mathcal{Q} $ when the node moves into the cluster of each of its neighbors. 2) Merge the node with the neighbor node that the maximum $\Delta \mathcal{Q} $ (only if $\Delta \mathcal{Q} >$ 0) obtained by the calculation. Repeat until the clusters balance each other (the maximum $\Delta \mathcal{Q} $ is less than 0). Thus, the nodes in $G$ that belong to the same cluster are aggregated into one super node, and obtain $G$${}' $.

\textbf{Node-Supernode:} For contrastive learning between node $x_{i}$ in $G$ and supernode $y_{j}$ in $G$${}' $, where $x_{i}$ is refine from $y_{j}$, we define a positive example as:
\begin{equation}
	\mathcal{X}^{'} =\left \{ \left ( x_{i} ,\textbf{s}_{j}  \right )  \mid x_{i}  \propto  y_{j}\right \}
\end{equation}
where $\textbf{s}_{j}$ is the midpoint of supernode $y_{j}$.
The supernode $y_{1}$ in network $G$${}' $ is thus regarded as a collection of nodes $x_{1}$, $x_{2}$,$\cdots$,$x_{n}$ in $G$. The nodes in the collection are the positive examples  of $\textbf{s}_{j}$.
Then,  the contrastive loss of a pair of positive nodes-supernodes in the network is as follows:
\begin{equation}
	l_{n2s}^{} \left ( v_{i} , y_{j}\right )  = - \log_{}{\frac{exp\left ( sim_{\kappa}\left ( x_{i} ,\textbf{s}_{i} \right )  \right ) }{\sum_{g=1}^{N^{'}}exp\left ( sim_{\kappa}\left ( x_{i} ,\textbf{s}_g \right ) \right ) }} 
\end{equation}
where $N^{'}$ is the number of supernodes in  $G$${}' $. The contrastive  loss of node-supernode is as follows:
\begin{equation}
	\mathcal{L}_{n2s}^{}   = \frac{1}{2N^{'}} \sum_{k=1}^{N^{'}} \frac{1}{\left | \mathcal{X}^{'}\left ( k \right )   \right | }\sum_{\left ( i,j \right )\in \mathcal{X}^{'} \left ( k \right )  }^{}\left ( l_{n2s}  \left ( v_{i},y_{j}  \right ) +  l _{n2s} \left ( y_{j},v_{i}  \right )\right ),   
\end{equation}
Consequently, the final intra-network contrastive loss is as follows: 
\begin{equation}
	\mathcal{L}_{intra}^{}   =\mathcal{L}_{in}+ \mathcal{L}_{n2s}.
\end{equation}


\noindent\textbf{Inter-network Contrastive Learning.}
For inter-contrastive learning, we highlight the anchor users and study the similarity among users of different network layers.
For each anchor, we maximize the representation agreement between user of one network and its counterpart of the other network in the common tangent space. 

For the two networks ${G}_{i}$ and ${G}_{j}$ that need to be aligned, we take the pre-known anchor link between the two networks as a positive example. 
Due to the large number of unknown aligned nodes between the two networks, we consider enhancing our model representation with unanchored linked nodes. For example, $v_{1}$ and $v_{1}'$ are at the same small granularity in ${G}_{1}$; $v_{2}$ and $v_{2}'$ are at the same small granularity in ${G}_{2}$. If there is an anchor link between these two small granularities, then there are also anchor links between $v_{1}$ and $v_{2}$, $v_{1}'$ and $v_{2}'$.

First, we define a pair of positive nodes: $\mathcal{D}=\left \{ (i,j)\mid \left [ \boldsymbol{D} \right ]_{ij} >0  \right \}  $, where $\boldsymbol{D}$ is the matrix of pre-anchor links between ${G}_{i}$ and ${G}_{j}$.
Then we can define the loss function between a pair of positive nodes $(v^{i}_{s},v^{j}_{t} )$ as:
\begin{equation}
	l_{er} \left ( v^{i}_{s} ,v^{j}_{t}  \right ) =-\log_{}{\frac{exp(sim_{\kappa}(x_{s}^{i} ,x_{t}^{j}))}{ {\textstyle \sum_{h=1}^{N}}   exp( sim_{\kappa}(x_{s}^{i} ,x_{t}^{h}))} } 
\end{equation}
where $sim_{\kappa}(x,y)=\left \langle log_{\nu }^{\kappa  } (x),log_{\nu }^{\kappa } (y) \right \rangle  $.
Finally, we define the loss between the two networks ${G}_{i}$ and ${G}_{j}$ as:
\begin{equation}
\mathcal{L}_{inter} =\frac{1}{2\left | \mathcal{D}  \right | } \sum_{(s,t)\in \mathcal{D} }\left ( l_{er}\left ( v_{s}^{i}  ,v_{t}^{j}   \right ) +l_{er}\left ( v_{t}^{j}, v_{s}^{i}  \right )   \right ) 
\end{equation}
\noindent\textbf{Overall Objective of RCoCo.}
Finally, the overall objective  is given by integrating intra- \& inter-contrastive loss and supernode loss,
\begin{equation}
	\mathcal{J}_{RCoCo} =\mathcal{L}_{intra}+ \mathcal{L}_{inter}+\alpha \mathcal{Q},
\end{equation}
where $\alpha$ is a weighting factor. The training procedure is summarized in Algorithm 1. 
\emph{Consequently, RCoCo learns the representation of each layer in the respective manifolds, who are aligned in the common tangent space on the anchors, so that the expressive representations are capable of conducting collective link prediction.}

\noindent\textbf{Computational Complexity.} Here, we specify the computational complexity of Algorithm \ref{algo_disjdecomp}.
\begin{itemize}
	\item \textit{I-Louvain}: The time complexity is $O(|\mathcal E|)$, where $\mathcal E$ is the edge set. The space complexity is $O(|\mathcal V|^{2})$, where $\mathcal V$ is the number of nodes in the graph.
	\item \textit{Intra-Contrastive Learning}: The time complexity of Node-to-Node and Node-to-Supernode contrast is $O(|\mathcal V|^2)$ and $O(K|\mathcal V|)$, respectively. $\mathcal V$ is the node set, and $K$ is the number of supernodes. The space complexity involves generating matrices A and X. The space complexity of matrix A is $O(|\mathcal V|^2)$, and the space complexity of matrix X is $O(|\mathcal V|d)$, where d is the feature dimension and $\mathcal V$ is the number of nodes. 
	\item  \textit{Intre-Contrastive Learning}: The time complexity of Inter-contrastive loss is $O(|\mathcal D| |\mathcal V|)$, where $\mathcal D$ is the anchor set. It involves A and X, and the space complexity is partially similar to Intra-Contrastive Learning.
\end{itemize}
Note that, Node-to-Node of Intra-Contrastive Learning is the most costly component, and thus the computational complexity is in the order of $O(|\mathcal V|^2)$.

\section{Experimental Setups}

In the experimental setups, we introduce the datasets, baselines, evaluation metrics and other implementation details.

\subsection{Datasets}

Without loss of generality, we select a variety of datasets: two social network pairs, FB-TW$_A$ \cite{Facebook-TwitterCaoY16}, TW$_B$-FS \cite{tkdeSunZWJWSY23}, a knowledge graph pair, {DBpedia}$_{\text{CH}}$-{DBpedia}$_{\text{EN}}$ \cite{semwebSunHL17} and an academic network pair, AMiner-DBLP \cite{kddArnetMiner08} \cite{tkdeSunZWJWSY23}. The statistics are shown in Table \ref{tab:t3}.

\begin{table}[]
	\centering
	\caption{Experimental datasets.}
	\label{tab:t3}
	\vspace{0.1in}
	\begin{tabular}{ccccc}
		\hline
		\multicolumn{2}{c}{Datasets}          & Nodes   & Links      & Align users              \\ \hline
		\multirow{2}{*}{FB-TW$_A$}    & Facebook   & 422,291 & 3,710,789  & \multirow{2}{*}{328,244} \\
		& TwitterA    & 669,198 & 12,749,257 &                          \\\hline
		\multirow{2}{*}{TW$_B$-FS}    & TwitterB    & 5,167  & 164,660    & \multirow{2}{*}{2,858}  \\
		& Foursquare & 5,240  & 76,972    &                          \\\hline
		\multirow{2}{*}{DBpedia}$_{\text{CH-EN}}$ & DBpedia$_{CH}$ & 66,469  & 153,929    & \multirow{2}{*}{42,540}        \\
		& DBpedia$_{EN}$ & 98,125  & 237,674    &                          \\\hline
		\multirow{2}{*}{AMiner-DBLP} & AMiner     & 26,386  & 273,476    & \multirow{2}{*}{18,255}  \\
		& DBLP       & 24,352  & 316,565    &                          \\ \hline
	\end{tabular}
\end{table}

\textbf{FB-TW$_A$.} 
Datasets of Facebook and Twitter$_A$ are collected by Cao et al. \cite{Facebook-TwitterCaoY16}, where the alignment information is given by  a third-party platform: About.Me. Facebook and Twitter are two popular social media platforms that allow users to create profiles, share photos and videos, send messages, and interact with friends.
In the network of Facebook/Twitter, nodes are the user accounts, and edges describe the friendship or following in the platform.

\textbf{TW$_B$-FS.} 
Foursquare is a location-based social platform that provides location data, business information, and recommendations. Users can share places they've been, check in, and interact with friends on it. Users are the nodes, and friending behavior forms the edges.
Note that, TW$_A$ and TW$_B$ are two different subsets in the network of Twitter.

\textbf{DBpedia}$_{\text{CH}}$-\textbf{DBpedia}$_{\text{EN}}$.  
DBpedia is a large-scale multilingual knowledge base. We choose the knowledge bases in two languages: Chinese and English, denoted as DBpedia$_{\text{CH}}$ and DBpedia$_{\text{EN}}$, respectively. For the tuples of \emph{(head entity, relation, end entity)} in the knowledge base, we regard entities as the nodes in the network, and relations as the edges. (We do not distinguish different types of relations in the induced network). The entity alignment between different languages is given in \cite{semwebSunHL17}.

\textbf{AMiner-DBLP.} 
Both AMiner and DBLP are academic collaboration networks. AMiner is an academic search engine. DBLP is a database of academic papers in the field of computer science. Both of them provide information such as papers, author profiles, publishers, etc. 
Note that, nodes are the researchers, and two nodes are linked if the researchers have coauthored  at least one academic paper.


\subsection{Learning Tasks \& Evaluation Metrics}

With the aim of intra-link prediction and inter-link prediction, we include two groups of baselines. 
For the former, we employ AUC and F1 as evaluation metrics, and choose $6$ strong baselines introduced as follows.

\begin{itemize}
\item \textbf{GCN} \cite{iclrGCN17}: GCN is a semi-supervised learning method on the graph to learn the representation via a local first-order approximation of  convolution. GCN stacks the building block layer, convolution layer, that aggregates the information in the neighborhood from the spectral perspective.


\item \textbf{GAT} \cite{iclrGAT18}: GAT generalize the attention mechanism to the graph domain. Different from GCN, the   building block layer is the attention layer which aggregates the information in the neighborhood from the special perspective.

\item \textbf{DGI} \cite{DGI}: DGI is a popular self-supervised graph learning method built upon the graph convolutional networks. It introduces the patch representations, which summarizes a subgraph centered around a certain node, for graph augmentation and maximize the mutual information to learn node representations.

\item \textbf{GCLN} {\cite{IEEEWU22}}: GCLN explores the interaction of attraction and repulsion on the graphs. The attraction encourages features from both graph domains to be consistent, and the repulsion ensures the differentiation of features, solving the limitations of cross-graph domains to a certain extent.

\item \textbf{Heco} {\cite{tkdeLiu23}}: Heco is a recent self-supervised learning method based on the co-contrast mechanism. It contrasts the network-mode view and a meta-path view to capture both the local and higher-order structure of the graph.

\item \textbf{HGCN} \cite{nipsChamiYRL19}: HGCN extends graph convolutional neural networks to the domain of hyperbolic geometry. It designs attentional convolution in hyperbolic geometry to learn the node structure embedding of the graph. Note that, it cannot work with the manifold beyond  hyperbolic geometry, e.g., hyperspherical spaces.
\end{itemize}

Note that, the popular GCN, GAT, DGI and the recent GCLN and Heco are designed in the traditional Euclidean space, and cannot be directly adapted to other manifold due to the inherent difference in geometry. To the best of our knowledge, there exists few works studying link prediction in generic manifold.

On inter-link prediction (network alignment), we choose $7$ strong baselines that are introduced as follows.


\begin{itemize}
\item \textbf{Moana} \cite{Moana19}: Moana proposes an algorithm for multi-layer network alignment using coarsen-alignment-interpolation framework. It first coarsely structures the input network, then aligns the coarse network, and then uses interpolation to align fine-grained nodes, enriching the multi-level input network alignment patterns.

\item \textbf{G-CREWE} \cite{Gcrewecorr}: G-CREWE  aligns the network with two levels of resolution: the fine resolution of the original network and the coarse resolution of the compressed network. This method leverages GCN to embed and compress networks, in order to facilitate alignment.

\item \textbf{SNNA} \cite{aaaiSNNA19}: SNNA considers network alignment at the distribution level of social networks. It models the social network as a whole, and learns the weakly-supervised, adversarial projection function of all user identities.

\item \textbf{Meta-NA} \cite{Meta}: Meta-NA follows the semi-supervised setting. It leverages the known anchor nodes for  meta-learning, acquires a unified metric space, and obtains the potential prior knowledge of unknown anchor links, which improves the generalization of anchor link nodes.

\item \textbf{IONE-D} \cite{IONE-D}: IONE-D leverages the social structure information, and encodes the follower/followee relationship as input and output vector of user representations, helping retain high similarity users for network alignment.

\item \textbf{NeXtAlign} \cite{NeXtAlign}: NeXtAlign achieves a balance between network alignment consistency and disparity, and quantifies the learning risk of sampling distribution based on the graph convolutional neural network.

\item \textbf{NAME} \cite{NAME}: NAME is an end-to-end alignment framework. It first creates node embeddings under different modes, and then design a late-fusion mechanism for embedding integration in account of mode importance.

\item \textbf{HUIL} \cite{IEEEWang20}: HUIL  introduces hyperbolic geometry to network alignment. It optimizes the joint objective of user embedding and alignment.

\end{itemize}

Note that, all the baselines except HUIL are Euclidean methods. HUIL assume different networks have the same curvature, different from our design. Recently, Sun et. al. \cite{DBLP:conf/icdm/0008Z0WDSY20} focus on network alignment at the community level in hyperbolic space, which is orthogonal to our study. 

We employ $Hit@K$ and $MRR@K$ as the evaluation metrics.
Specifically, for each node $v_{i}^{s}  \in \mathcal{V }^{s}$, we obtain a candidate list, which consists of possible anchor nodes in $\mathcal{V}^{s}$, and select $k$ from the top as the candidate list.
\begin{equation}
	hit\left ( k \right ) =\frac{1}{\mathcal{D} } \sum_{\left ( i,j \right ) \in \mathcal{D} }^{} \left ( success\left ( k \right )  \right ) 
\end{equation}
where $success(k)$ indicates whether $v_{j}^{t}$ is in the $v_{i}^{s}$ candidate list of length $k$.
\begin{equation}
	MRR=\frac{1}{n} \sum_{\left ( i,j \right )\in \mathcal{D}  }^{} \frac{1}{rank_{i} } 
\end{equation}
$rank_{i}$ represents the rank of $v_{j}^{t}$ in the candidate list of $v_{i}^{s}$, and if it does not exist, $rank_{i} \to \infty$ ($\frac{1}{rank_{i}}=0$).

\subsection{Model Configuration}

Here, we detail model configuration to enhance reproducibility. In RCoCo, we stack the $\kappa$-attentional convolution twice (the number of hidden layers is 1), and the dropout is 0.3 by default. On the loss, the weighting factor $\alpha$ is set as 10, as accurate supernode is important to the contrastive learning in the manifold. The number of supernodes is set in accordance to the number of classes in the datasets. The representation dimension is set as $32$ by default. Note that, dimensions of the baselines are set according to the original papers, and we will further discuss on this issue in Sec. 5.4.
The manifold-valued representations are optimized by Riemannian Adam, while the others are optimized by Adam. The learning rate of both optimizer is set as 0.001, and the batch size is 3000. For user alignment, the hyperparameter of network overlap rate is set as $20\%$ by default.

\subsection{Hardware \& Software}

All experiments in our paper are done on an NVIDIA Tesla V100 GPU.
Our model is built upon PyTorch Geometric and GeoOpt \cite{icmlcurvatureGCN20}. 

\section{Results and Discussion}

We conduct extensive experiments with $14$ strong baselines on $8$ real-world networks, leading by the research questions (RQ) as follows:
\begin{itemize}
	\item RQ1: How does RCoCo perform for intra-link prediction? 
	\item RQ2: How does RCoCo perform for inter-link prediction (network alignment)?  
	\item RQ3: How does the designs in our solution contribute to the success of RCoCo? 
	\item RQ4: How is the parameter sensitivity of  RCoCo? 
\end{itemize}
For fair comparison, we repeat each baselines $10$ times independently, and report the mean performance as well as standard derivation in this section.

\subsection{Main Results I: Intra-network Link Prediction}

We compare RCoCo with GCN, GAT, DGI, GCLN, Heco, HGCN and $\kappa-$GAT, a proposed component of RCoCo. 
Concretely, $\kappa-$GAT is trained with the intra-contrastive loss, and the inclusion of $\kappa-$GAT  is to evaluate the effectiveness of curvature adaptation.
Note that, for RCoCo, we feed the multiplex network into our model, and obtain predicted intra-links of all the layers.
For the cases that only the data of a single network is available, RCoCo is degraded as the $\kappa-$GAT accordingly. This is another reason why we evaluate the performance of $\kappa-$GAT.
With user representations learnt, we leverage the Fermi-Dirac decoder to predict the intra-links. The distance metric of the decoder is set with respect to the manifold of comparison methods. For instance, we employ the generic distance $d_\kappa$ for RCoCo in arbitrary CCS, and hyperbolic distance for HGCN.
We examine the performance on all the $6$ datasets. For each dataset, we split training set, validation set and testing set according to the ratio of $85\%-5\%-15\%$.
Empirical results in terms of AUC and F1 are recorded in Table \ref{tab:t6} and \ref{tab:t7}, respectively. 
As shown in the tables, the proposed RCoCo consistently achieves the best results for all the cases. The reason is two-fold: 1) RCoCo collaborates the intra- \& inter-network behaviors, so that the information of counterpart network facilitates intra-link prediction in the target network. 2) RCoCo learns the inherent geometry of network structures via the curvature estimator, strengthening the expressiveness  and lowering the distortion of user representations.
Surprisingly, $\kappa-$GAT is able to achieve competitive results to RCoCo, and outperforms most of the well-designed baselines.
Also, we notice that HGCN performs well on the FB, FS, DBLP, AMiner, TW$_A$ and TW$_B$ datasets of low $\delta-$hyperbolicity, recalling the empirical investigation in Table 1.
Both observations verify the importance of the geometric bias that the representation space needs to match the geometry of the data. (We will further discuss the issue of geometric bias in the Sec. 5.3 of Ablation Study.) In the meanwhile, it demonstrates the effectiveness of our plug-and-use curvature estimator.

\begin{table*}[]
	\centering
	\caption{Intra-link prediction results on FB, TW$_A$, TW$_B$, FS, DBpedia$_{\text{CH}}$, DBpedia$_{\text{EN}}$, AMiner and DBLP datasets in terms of AUC (\%), where standard derivations are given in brackets. The best results are in the \textbf{boldfaced}, and the second are \underline{underlined}.}
	\vspace{0.1in}
	\label{tab:t6}
	\resizebox{\linewidth}{!}{
		\begin{tabular}{c|c|c|c|c|c|c|c|c}
			\hline
			\diagbox[width=6.5em]{Method}{Dataset} & FB & TW$_A$ & TW$_B$ & FS & DB$_\text{CH}$ & DB$_\text{EN}$ & AMiner & DBLP \\ \hline
			GCN     &\begin{tabular}[c]{@{}c@{}}85.87\\ \scriptsize{($\pm $0.47)}\end{tabular}       & \begin{tabular}[c]{@{}c@{}}86.13\\ \scriptsize{($\pm $0.12)}\end{tabular}        &\begin{tabular}[c]{@{}c@{}}85.97\\ \scriptsize{($\pm $0.19)}\end{tabular}         &\begin{tabular}[c]{@{}c@{}}83.96\\ \scriptsize{($\pm $0.28)}\end{tabular}     &\begin{tabular}[c]{@{}c@{}}87.05\\ \scriptsize{($\pm $0.11)}\end{tabular}                 & \begin{tabular}[c]{@{}c@{}}87.21\\ \scriptsize{($\pm $0.05)}\end{tabular}                &\begin{tabular}[c]{@{}c@{}}88.95\\ \scriptsize{($\pm $0.21)}\end{tabular}         & \begin{tabular}[c]{@{}c@{}}89.63\\ \scriptsize{($\pm $0.31)}\end{tabular}      \\ 
			GAT     &\begin{tabular}[c]{@{}c@{}}89.32\\ \scriptsize{($\pm $0.25)}\end{tabular}    &\begin{tabular}[c]{@{}c@{}}88.95\\ \scriptsize{($\pm $0.25)}\end{tabular}        &\begin{tabular}[c]{@{}c@{}}89.13\\ \scriptsize{($\pm $0.28)}\end{tabular}        &\begin{tabular}[c]{@{}c@{}}88.15\\ \scriptsize{($\pm $0.23)}\end{tabular}    & \begin{tabular}[c]{@{}c@{}}86.93\\ \scriptsize{($\pm $0.51)}\end{tabular}               & \begin{tabular}[c]{@{}c@{}}87.11\\ \scriptsize{($\pm $0.48)}\end{tabular}               &  \begin{tabular}[c]{@{}c@{}}92.12\\ \scriptsize{($\pm $0.04)}\end{tabular}      & \begin{tabular}[c]{@{}c@{}}92.55\\ \scriptsize{($\pm $0.16)}\end{tabular}     \\
			
			DGI     &\begin{tabular}[c]{@{}c@{}}86.25\\ \scriptsize{($\pm $0.40)}\end{tabular}    &\begin{tabular}[c]{@{}c@{}}86.77\\ \scriptsize{($\pm $0.09)}\end{tabular}        & \begin{tabular}[c]{@{}c@{}}88.50\\ \scriptsize{($\pm $0.04)}\end{tabular}       &\begin{tabular}[c]{@{}c@{}}87.02\\ \scriptsize{($\pm $0.01)}\end{tabular}    
			&\begin{tabular}[c]{@{}c@{}}88.23\\ \scriptsize{($\pm $0.07)}\end{tabular}                &\begin{tabular}[c]{@{}c@{}}88.56\\ \scriptsize{($\pm $0.01)}\end{tabular}                &\begin{tabular}[c]{@{}c@{}}90.05\\ \scriptsize{($\pm $0.12)}\end{tabular}        &\begin{tabular}[c]{@{}c@{}}90.26\\ \scriptsize{($\pm $0.10)}\end{tabular}      \\

			GCLN    &\begin{tabular}[c]{@{}c@{}}91.57\\ \scriptsize{($\pm $0.06)}\end{tabular}    & \begin{tabular}[c]{@{}c@{}}91.86\\ \scriptsize{($\pm $0.40)}\end{tabular}       & \begin{tabular}[c]{@{}c@{}}91.15\\ \scriptsize{($\pm $0.03)}\end{tabular}       & \begin{tabular}[c]{@{}c@{}}90.22\\ \scriptsize{($\pm $0.17)}\end{tabular}   & \begin{tabular}[c]{@{}c@{}}86.70\\ \scriptsize{($\pm $0.20)}\end{tabular}               &  \begin{tabular}[c]{@{}c@{}}87.25\\ \scriptsize{($\pm $0.39)}\end{tabular}              &\begin{tabular}[c]{@{}c@{}}92.11\\ \scriptsize{($\pm $0.06)}\end{tabular}        & \begin{tabular}[c]{@{}c@{}}91.90\\ \scriptsize{($\pm $0.22)}\end{tabular}     \\ 
			
			Heco    &\begin{tabular}[c]{@{}c@{}}87.95\\ \scriptsize{($\pm $0.12)}\end{tabular}    & \begin{tabular}[c]{@{}c@{}}88.03\\ \scriptsize{($\pm $0.31)}\end{tabular}       &\begin{tabular}[c]{@{}c@{}}87.92\\ \scriptsize{($\pm $0.27)}\end{tabular}        & \begin{tabular}[c]{@{}c@{}}88.26\\ \scriptsize{($\pm $0.10)}\end{tabular}   & \begin{tabular}[c]{@{}c@{}}88.16\\ \scriptsize{($\pm $0.27)}\end{tabular}               & \begin{tabular}[c]{@{}c@{}}88.30\\ \scriptsize{($\pm $0.41)}\end{tabular}               &\begin{tabular}[c]{@{}c@{}}91.07\\ \scriptsize{($\pm $0.09)}\end{tabular}        & \begin{tabular}[c]{@{}c@{}}91.33\\ \scriptsize{($\pm $0.32)}\end{tabular}     \\ 
			
			HGCN    &\begin{tabular}[c]{@{}c@{}}$\underline{91.99}$\\ \scriptsize{($\pm $0.07)}\end{tabular}    & \begin{tabular}[c]{@{}c@{}}\underline{92.01}\\ \scriptsize{($\pm $0.23)}\end{tabular}       & \begin{tabular}[c]{@{}c@{}}\underline{91.87}\\ \scriptsize{($\pm $0.39)}\end{tabular}       & \begin{tabular}[c]{@{}c@{}}\underline{91.15}\\ \scriptsize{($\pm $0.41)}\end{tabular}   & \begin{tabular}[c]{@{}c@{}}86.07\\ \scriptsize{($\pm $0.39)}\end{tabular}               & \begin{tabular}[c]{@{}c@{}}86.92\\ \scriptsize{($\pm $0.43)}\end{tabular}               & \begin{tabular}[c]{@{}c@{}}\underline{92.99}\\ \scriptsize{($\pm $0.12)}\end{tabular}       & \begin{tabular}[c]{@{}c@{}}\underline{93.07}\\ \scriptsize{($\pm $0.08)}\end{tabular}     \\

			$\kappa$-GAT   &\begin{tabular}[c]{@{}c@{}}91.67\\ \scriptsize{($\pm $0.42)}\end{tabular}    & \begin{tabular}[c]{@{}c@{}}91.89\\ \scriptsize{($\pm $0.17)}\end{tabular}       & \begin{tabular}[c]{@{}c@{}}91.85\\ \scriptsize{($\pm $0.48)}\end{tabular}       &\begin{tabular}[c]{@{}c@{}}91.09\\ \scriptsize{($\pm $0.16)}\end{tabular}    & \begin{tabular}[c]{@{}c@{}}\underline{90.13}\\ \scriptsize{($\pm $0.04)}\end{tabular}               & \begin{tabular}[c]{@{}c@{}}\underline{90.55}\\ \scriptsize{($\pm $0.05)}\end{tabular}               & \begin{tabular}[c]{@{}c@{}}92.81\\ \scriptsize{($\pm $0.21)}\end{tabular}       & \begin{tabular}[c]{@{}c@{}}92.89\\ \scriptsize{($\pm $0.11)}\end{tabular}     \\ \hline

			\textbf{RCoCo}   
			& \begin{tabular}[c]{@{}c@{}}\textbf{92.24}\\ \scriptsize{{($\pm $0.11)}}\end{tabular}    
			&\begin{tabular}[c]{@{}c@{}}\textbf{92.63}\\ \scriptsize{{($\pm $0.20)}}\end{tabular}         & \begin{tabular}[c]{@{}c@{}}\textbf{92.16}\\ \scriptsize{{($\pm $0.07)}}\end{tabular}        & \begin{tabular}[c]{@{}c@{}}\textbf{91.59}\\ \scriptsize{{($\pm $0.25)}}\end{tabular}    & \begin{tabular}[c]{@{}c@{}}\textbf{90.76}\\ \scriptsize{{($\pm $0.22)}}\end{tabular}                &\begin{tabular}[c]{@{}c@{}}\textbf{90.69}\\ \scriptsize{{($\pm $0.20)}}\end{tabular}               
			 & \begin{tabular}[c]{@{}c@{}}\textbf{93.25}\\ \scriptsize{{($\pm $0.03)}}\end{tabular}        & \begin{tabular}[c]{@{}c@{}}\textbf{93.97}\\ \scriptsize{{($\pm $0.20)}}\end{tabular}      \\ \hline
	\end{tabular}}
\end{table*}

\begin{table*}[]
	\centering
	\caption{Intra-link prediction results on FB, TW$_A$, TW$_B$, FS, DBpedia$_{\text{CH}}$, DBpedia$_{\text{EN}}$, AMiner and DBLP datasets in terms of F1 (\%), where standard derivations are given in brackets. The best results are in the \textbf{boldfaced}, and the second are \underline{underlined}.}
	\vspace{0.1in}
	\label{tab:t7}
	\resizebox{\linewidth}{!}
	{
		\begin{tabular}{c|c|c|c|c|c|c|c|c}
			\hline
			\diagbox[width=6.5em]{Method}{Dataset} & FB & TW$_A$ & TW$_B$ & FS & DB$_\text{CH}$ & DB$_\text{EN}$ & AMiner & DBLP \\ \hline
			GCN     &\begin{tabular}[c]{@{}c@{}}86.11\\ \scriptsize{($\pm $0.20)}\end{tabular}    &\begin{tabular}[c]{@{}c@{}}86.58\\ \scriptsize{($\pm $0.32)}\end{tabular}        &\begin{tabular}[c]{@{}c@{}}86.26\\ \scriptsize{($\pm $0.31)}\end{tabular}        &\begin{tabular}[c]{@{}c@{}}83.96\\ \scriptsize{($\pm $0.13)}\end{tabular}    &\begin{tabular}[c]{@{}c@{}}88.12\\ \scriptsize{($\pm $0.51)}\end{tabular}               
			&\begin{tabular}[c]{@{}c@{}}87.93\\ \scriptsize{($\pm $0.30)}\end{tabular}               
			&\begin{tabular}[c]{@{}c@{}}89.16\\ \scriptsize{($\pm $0.04)}\end{tabular}       
			&\begin{tabular}[c]{@{}c@{}}90.13\\ \scriptsize{($\pm $0.11)}\end{tabular}     \\

			GAT     &\begin{tabular}[c]{@{}c@{}}90.05\\ \scriptsize{($\pm $0.21)}\end{tabular}    & \begin{tabular}[c]{@{}c@{}}88.70\\ \scriptsize{($\pm $0.18)}\end{tabular}     &
			\begin{tabular}[c]{@{}c@{}}89.91\\ \scriptsize{($\pm $0.29)}\end{tabular}        &
			\begin{tabular}[c]{@{}c@{}}88.15\\ \scriptsize{($\pm $0.08)}\end{tabular}   & \begin{tabular}[c]{@{}c@{}}87.05\\ \scriptsize{($\pm $0.42)}\end{tabular}               &
			\begin{tabular}[c]{@{}c@{}}87.24\\ \scriptsize{($\pm $0.35)}\end{tabular}                &
			\begin{tabular}[c]{@{}c@{}}92.67\\ \scriptsize{($\pm $0.13)}\end{tabular}       & \begin{tabular}[c]{@{}c@{}}92.93\\ \scriptsize{($\pm $0.23)}\end{tabular}     \\

			DGI     &\begin{tabular}[c]{@{}c@{}}88.75\\ \scriptsize{($\pm $0.22)}\end{tabular}    &
			\begin{tabular}[c]{@{}c@{}}87.11\\ \scriptsize{($\pm $0.41)}\end{tabular}        & \begin{tabular}[c]{@{}c@{}}89.02\\ \scriptsize{($\pm $0.10)}\end{tabular}       &
			\begin{tabular}[c]{@{}c@{}}87.02\\ \scriptsize{($\pm $0.10)}\end{tabular}        & \begin{tabular}[c]{@{}c@{}}89.20\\ \scriptsize{($\pm $0.22)}\end{tabular}   & \begin{tabular}[c]{@{}c@{}}88.93\\ \scriptsize{($\pm $0.11)}\end{tabular}               & \begin{tabular}[c]{@{}c@{}}89.99\\ \scriptsize{($\pm $0.40)}\end{tabular}               &
			\begin{tabular}[c]{@{}c@{}}90.35\\ \scriptsize{($\pm $0.15)}\end{tabular}     \\

			GCLN    &\begin{tabular}[c]{@{}c@{}}91.68\\ \scriptsize{($\pm $0.10)}\end{tabular}    &\begin{tabular}[c]{@{}c@{}}92.05\\ \scriptsize{($\pm $0.27)}\end{tabular}        
			&\begin{tabular}[c]{@{}c@{}}91.72\\ \scriptsize{($\pm $0.15)}\end{tabular}       &\begin{tabular}[c]{@{}c@{}}90.22\\ \scriptsize{($\pm $0.21)}\end{tabular}   &\begin{tabular}[c]{@{}c@{}}87.91\\ \scriptsize{($\pm $0.05)}\end{tabular}               &\begin{tabular}[c]{@{}c@{}}87.67\\ \scriptsize{($\pm $0.06)}\end{tabular}                &\begin{tabular}[c]{@{}c@{}}92.31\\ \scriptsize{($\pm $0.14)}\end{tabular}       
			&\begin{tabular}[c]{@{}c@{}}92.11\\ \scriptsize{($\pm $0.27)}\end{tabular}     \\

			Heco    &\begin{tabular}[c]{@{}c@{}}88.20\\ \scriptsize{($\pm $0.12)}\end{tabular}    
			&\begin{tabular}[c]{@{}c@{}}89.31\\ \scriptsize{($\pm $0.18)}\end{tabular}       &\begin{tabular}[c]{@{}c@{}}88.65\\ \scriptsize{($\pm $0.23)}\end{tabular}        
			&\begin{tabular}[c]{@{}c@{}}88.26\\ \scriptsize{($\pm $0.15)}\end{tabular}   &\begin{tabular}[c]{@{}c@{}}88.16\\ \scriptsize{($\pm $0.13)}\end{tabular}                &\begin{tabular}[c]{@{}c@{}}87.56\\ \scriptsize{($\pm $0.15)}\end{tabular}              
			&\begin{tabular}[c]{@{}c@{}}91.52\\ \scriptsize{($\pm $0.46)}\end{tabular}       
			&\begin{tabular}[c]{@{}c@{}}91.51\\ \scriptsize{($\pm $0.26)}\end{tabular}     \\

			HGCN    &\begin{tabular}[c]{@{}c@{}}$\underline{92.13}$\\ \scriptsize{($\pm $0.17)}\end{tabular}   & \begin{tabular}[c]{@{}c@{}}92.20\\ \scriptsize{($\pm $0.29)}\end{tabular}       & \begin{tabular}[c]{@{}c@{}}92.03\\ \scriptsize{($\pm $0.11)}\end{tabular}       &  \begin{tabular}[c]{@{}c@{}}\underline{92.35}\\ \scriptsize{($\pm $0.46)}\end{tabular}  &
			\begin{tabular}[c]{@{}c@{}}87.01\\ \scriptsize{($\pm $0.20)}\end{tabular}                &
			\begin{tabular}[c]{@{}c@{}}86.98\\ \scriptsize{($\pm $0.16)}\end{tabular}              & \begin{tabular}[c]{@{}c@{}}\underline{93.27}\\ \scriptsize{($\pm $0.37)}\end{tabular}       & \begin{tabular}[c]{@{}c@{}}\underline{93.55}\\ \scriptsize{($\pm $0.10)}\end{tabular}     \\

			$\kappa$-GAT  &\begin{tabular}[c]{@{}c@{}}92.11\\ \scriptsize{($\pm $0.22)}\end{tabular}   
			&
			\begin{tabular}[c]{@{}c@{}}\underline{92.21}\\ \scriptsize{($\pm $0.36)}\end{tabular}       &
			\begin{tabular}[c]{@{}c@{}}\underline{92.07}\\ \scriptsize{($\pm $0.23)}\end{tabular}        & \begin{tabular}[c]{@{}c@{}}92.33\\ \scriptsize{($\pm $0.21)}\end{tabular}   & \begin{tabular}[c]{@{}c@{}}\underline{89.89}\\ \scriptsize{($\pm $0.29)}\end{tabular}               & \begin{tabular}[c]{@{}c@{}}\underline{89.35}\\ \scriptsize{($\pm $0.21)}\end{tabular}               & \begin{tabular}[c]{@{}c@{}}93.22\\ \scriptsize{($\pm $0.24)}\end{tabular}       & \begin{tabular}[c]{@{}c@{}}93.03\\ \scriptsize{($\pm $0.24)}\end{tabular}     \\ \hline

			\textbf{RCoCo}   &\begin{tabular}[c]{@{}c@{}}\textbf{93.02}\\ \scriptsize{($\pm $\textbf{0.10})}\end{tabular}  
			 & \begin{tabular}[c]{@{}c@{}}\textbf{92.97}\\ \scriptsize{($\pm $\textbf{0.28})}\end{tabular}      
			 & \begin{tabular}[c]{@{}c@{}}\textbf{93.26}\\ \scriptsize{($\pm $\textbf{0.10})}\end{tabular}       
			 & \begin{tabular}[c]{@{}c@{}}\textbf{92.86}\\ \scriptsize{($\pm $\textbf{0.17})}\end{tabular}   
			 & \begin{tabular}[c]{@{}c@{}}\textbf{91.33}\\ \scriptsize{($\pm $\textbf{0.13})}\end{tabular}             
			 & \begin{tabular}[c]{@{}c@{}}\textbf{91.10}\\ \scriptsize{($\pm $\textbf{0.25})}\end{tabular}               
			 & \begin{tabular}[c]{@{}c@{}}\textbf{94.05}\\ \scriptsize{($\pm $\textbf{0.29})}\end{tabular}      
			 & \begin{tabular}[c]{@{}c@{}}\textbf{94.45}\\ \scriptsize{($\pm $\textbf{0.31})}\end{tabular}    \\ \hline
		\end{tabular}}
\end{table*}

\subsection{Main Results II: Inter-network Link Prediction}
The intra-link prediction results in terms of $Hit@K$ and $MRR@K$ are collected in Table X and Table X, respectively. Without loss of generality, we report two cases of $K=10$ and $K=15$. For each dataset,  we segment anchor users with $65\%-10\%-25\%$ split for training, validation and testing. Note that, in RCoCo, the alignment is performed in the common tangent space, which is Euclidean in the tight tangential domain, and thus anchor users are identified according to Euclidean distance.
As shown in the tables, HUIL obtains unsatisfactory results on DBpedia$_{\text{CH}}$-DBpedia$_{\text{EN}}$ network pair in contrast to the good result on the other network pairs. Recall the hyperspherical nature of DBpedia$_{\text{CH}}$ and DBpedia$_{\text{EN}}$. It suggests that performance loss tends to occur when the geometric bias is not in line with the data.
The proposed RCoCo consistently outperforms all of the competitors. In RCoCo, the joint optimization with intra-link prediction boosts inter-link prediction. More topological information is important to identify the (structural) invariance among different layers (e.g., common friending pattern), and thus facilitates to identify the anchor users.
Indeed, the observation of intra- \& inter-network collaboration motivates our study.

Furthermore, we study network alignment under scarce anchor annotations. In this case, we assume only a handful of anchors are known in prior, and set the training-validation-testing split as $5\%-10\%-85\%$ accordingly.
We show the alignment results on AMiner-DBLP and DBpedia$_{\text{CH}}$-DBpedia$_{\text{EN}}$  datasets in terms of $MRR@10$ in Fig. \ref{Fig.3}.
RCoCo achieves large MRR gain compared to other baselines.
The reason is that, the proposed contrastive loss of RCoCo 1) effectively learns informative user representations from the network itself, and 2) effectively aligns the manifolds of different layers in the common tangent space for alignment. 
Also, an appropriate manifold tends to alleviate this issue, as evidence in the results of HUIL on AMiner-DBLP dataset.

	\textbf{On the efficiency.} We have specified the computational complexity in the methodology, and shown that RCoCo has competitive complexity to other contrastive models. 	
	
\begin{figure*}[htbp]
	\centering 
	\includegraphics[width=13cm,height=4.0cm]{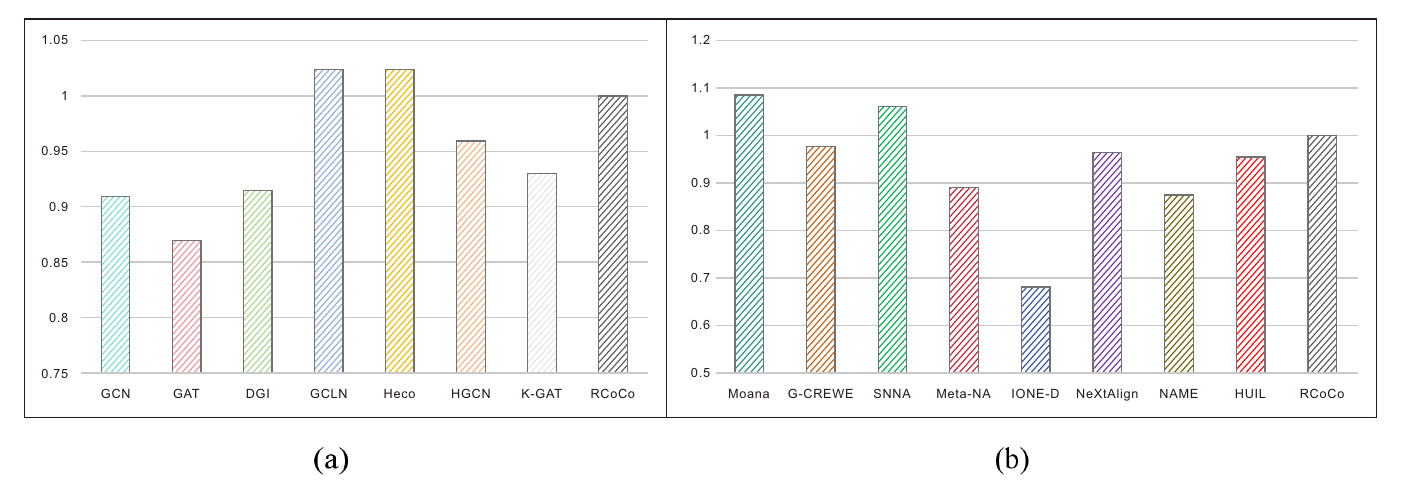}
	\caption{Efficiency of baselines and RCoCo. Taking the running time of RCoCo as the unit time, $(a)$ is the efficiency of link-prediction baselines relative to RCoCo, $(b)$ is the efficiency of user alignment baselines relative to RCoCo.
	}
	\label{Fig.23}
\end{figure*}

\begin{figure*}[htbp]
	\centering
	\includegraphics[width=13cm,height=4.5cm]{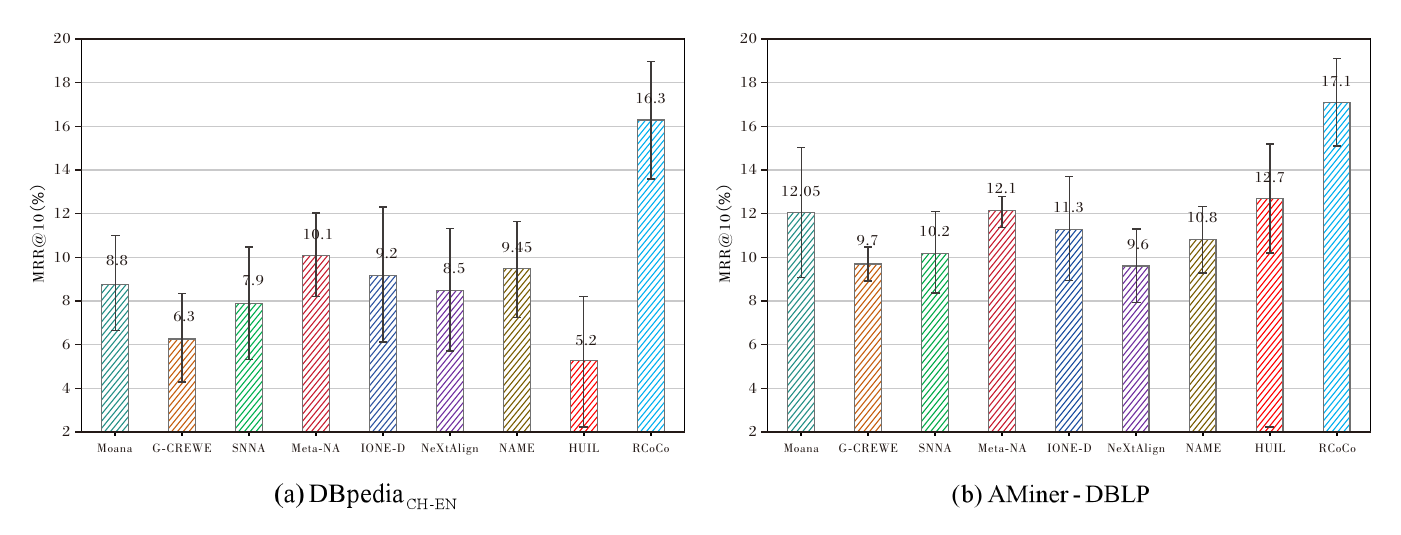}
	\caption{Alignment results on DBpedia$_{\text{CH}}$-DBpedia$_{\text{EN}}$ and AMiner-DBLP  datasets in terms of $MRR@10$.}
	\label{Fig.3}
\end{figure*}

	Here, we study the efficiency of RCoCo by evaluating the running time in practice. We summarize the running time of the baselines in  Fig. \ref{Fig.23}, where we use the running time of RCoCo  as the time unit. 
	As shown in Fig. \ref{Fig.23}, the running time of RCoCo is competitive to that of the state-of-the-art methods, given that Meta-NA of reinforcement learning is harder to converge than our RCoCo. Note that, RCoCo consistently achieves the best results in both intra- and inter-link prediction on all the datasets. In other words, the proposed model increases the accuracy without loss of efficiency.

\begin{table*}[]
	\centering
	\caption{Inter-link prediction results on FB-TW$_A$, TW$_B$-FS, DBpedia$_\text{CH-EN}$, AMiner-DBLP datasets in terms of $Hit@K$ (\%), where standard derivations are given in brackets. The best results are in the \textbf{boldfaced}, and the second are \underline{underlined}.}
	\vspace{0.1in}
	\label{tab:t4}
	\resizebox{\linewidth}{!}{
		\begin{tabular}{c|cc|cc|cc|cc}
			\hline
			{\multirow{2}*{\diagbox[width=6.5em]{Method}{Dataset}}} & \multicolumn{2}{c|}{FB-TW$_A$} & \multicolumn{2}{c|}{TW$_B$-FS} & \multicolumn{2}{c|}{DBpedia$_\text{CH-EN}$} & \multicolumn{2}{c}{AMiner-DBLP} \\
			& $k$=10           & $k$=15          & $k$=10           & $k$=15          & $k$=10                 & $k$=15                 & $k$=10           & $k$=15           \\ \hline
			Moana          & \begin{tabular}[c]{@{}c@{}}32.85\\ \scriptsize{($\pm $0.18)}\end{tabular}           & \begin{tabular}[c]{@{}c@{}}44.24\\ \scriptsize{($\pm $0.35)}\end{tabular}            &\begin{tabular}[c]{@{}c@{}}33.11\\ \scriptsize{($\pm $0.06)}\end{tabular}                      &\begin{tabular}[c]{@{}c@{}}41.56\\ \scriptsize{($\pm $0.22)}\end{tabular}                       & \begin{tabular}[c]{@{}c@{}}31.04\\ \scriptsize{($\pm $0.09)}\end{tabular}           & \begin{tabular}[c]{@{}c@{}}49.91\\ \scriptsize{($\pm $0.20)}\end{tabular}            &\begin{tabular}[c]{@{}c@{}}35.21\\ \scriptsize{($\pm $0.51)}\end{tabular}                      &\begin{tabular}[c]{@{}c@{}}42.05\\ \scriptsize{($\pm $0.11)}\end{tabular}                       \\
			G-CREWE        & \begin{tabular}[c]{@{}c@{}}26.94\\ \scriptsize{($\pm $0.33)}\end{tabular}            & \begin{tabular}[c]{@{}c@{}}36.39\\ \scriptsize{($\pm $0.27)}\end{tabular}             & \begin{tabular}[c]{@{}c@{}}29.13\\ \scriptsize{($\pm $0.12)}\end{tabular}                     & \begin{tabular}[c]{@{}c@{}}39.05\\ \scriptsize{($\pm $0.30)}\end{tabular}                      & \begin{tabular}[c]{@{}c@{}}28.41\\ \scriptsize{($\pm $0.26)}\end{tabular}            & \begin{tabular}[c]{@{}c@{}}38.54\\ \scriptsize{($\pm $0.06)}\end{tabular}             & \begin{tabular}[c]{@{}c@{}}31.30\\ \scriptsize{($\pm $0.07)}\end{tabular}                     & \begin{tabular}[c]{@{}c@{}}39.88\\ \scriptsize{($\pm $0.20)}\end{tabular}                     \\ 
			SNNA           & \begin{tabular}[c]{@{}c@{}}38.19\\ \scriptsize{($\pm $0.16)}\end{tabular}            &  \begin{tabular}[c]{@{}c@{}}52.37\\ \scriptsize{($\pm $0.18)}\end{tabular}            & \begin{tabular}[c]{@{}c@{}}35.05\\ \scriptsize{($\pm $0.10)}\end{tabular}                     & \begin{tabular}[c]{@{}c@{}}50.33\\ \scriptsize{($\pm $0.25)}\end{tabular}                      &  \begin{tabular}[c]{@{}c@{}}35.71\\ \scriptsize{($\pm $0.09)}\end{tabular}            &  \begin{tabular}[c]{@{}c@{}}54.84\\ \scriptsize{($\pm $0.05)}\end{tabular}             & \begin{tabular}[c]{@{}c@{}}33.75\\ \scriptsize{($\pm $0.23)}\end{tabular}                     & \begin{tabular}[c]{@{}c@{}}52.13\\ \scriptsize{($\pm $0.05)}\end{tabular}                     \\ 
			Meta-NA        &  \begin{tabular}[c]{@{}c@{}}29.89\\ \scriptsize{($\pm $0.15)}\end{tabular}            &  \begin{tabular}[c]{@{}c@{}}40.29\\ \scriptsize{($\pm $0.05)}\end{tabular}             & \begin{tabular}[c]{@{}c@{}}31.32\\ \scriptsize{($\pm $0.25)}\end{tabular}                     & \begin{tabular}[c]{@{}c@{}}44.28\\ \scriptsize{($\pm $0.11)}\end{tabular}                      &  \begin{tabular}[c]{@{}c@{}}25.37\\ \scriptsize{($\pm $0.28)}\end{tabular}            &  \begin{tabular}[c]{@{}c@{}}39.38\\ \scriptsize{($\pm $0.24)}\end{tabular}             & \begin{tabular}[c]{@{}c@{}}36.02\\ \scriptsize{($\pm $0.10)}\end{tabular}                     & \begin{tabular}[c]{@{}c@{}}49.60\\ \scriptsize{($\pm $1.08)}\end{tabular}                     \\ 
			IONE-D         &  \begin{tabular}[c]{@{}c@{}}41.08\\ \scriptsize{($\pm $0.27)}\end{tabular}            &  \begin{tabular}[c]{@{}c@{}}61.53\\ \scriptsize{($\pm $0.16)}\end{tabular}             & \begin{tabular}[c]{@{}c@{}}39.25\\ \scriptsize{($\pm $0.33)}\end{tabular}                     & \begin{tabular}[c]{@{}c@{}}59.16\\ \scriptsize{($\pm $0.08)}\end{tabular}                      &  \begin{tabular}[c]{@{}c@{}}37.31\\ \scriptsize{($\pm $0.13)}\end{tabular}            &  \begin{tabular}[c]{@{}c@{}}59.99\\ \scriptsize{($\pm $0.30)}\end{tabular}             & \begin{tabular}[c]{@{}c@{}}42.67\\ \scriptsize{($\pm $0.33)}\end{tabular}                     & \begin{tabular}[c]{@{}c@{}}61.24\\ \scriptsize{($\pm $0.24)}\end{tabular}                     \\ 
			NeXtAlign      &  \begin{tabular}[c]{@{}c@{}}\underline{43.88}\\ \scriptsize{($\pm $0.14)}\end{tabular}           &  \begin{tabular}[c]{@{}c@{}}\underline{64.27}\\ \scriptsize{($\pm $0.19)}\end{tabular}            & \begin{tabular}[c]{@{}c@{}}\underline{45.10}\\ \scriptsize{($\pm $0.20)}\end{tabular}                     & \begin{tabular}[c]{@{}c@{}}\underline{62.35}\\ \scriptsize{($\pm $0.15)}\end{tabular}                      &  \begin{tabular}[c]{@{}c@{}}\underline{40.16}\\ \scriptsize{($\pm $0.32)}\end{tabular}            &  \begin{tabular}[c]{@{}c@{}}\underline{63.47}\\ \scriptsize{($\pm $0.28)}\end{tabular}             & \begin{tabular}[c]{@{}c@{}}\underline{47.35}\\ \scriptsize{($\pm $0.67)}\end{tabular}                     & \begin{tabular}[c]{@{}c@{}}63.02\\ \scriptsize{($\pm $0.91)}\end{tabular}                     \\ 
			NAME           &  \begin{tabular}[c]{@{}c@{}}42.08\\ \scriptsize{($\pm $0.21)}\end{tabular}            &  \begin{tabular}[c]{@{}c@{}}63.66\\ \scriptsize{($\pm $0.10)}\end{tabular}             & \begin{tabular}[c]{@{}c@{}}40.67\\ \scriptsize{($\pm $0.18)}\end{tabular}                     & \begin{tabular}[c]{@{}c@{}}60.08\\ \scriptsize{($\pm $0.21)}\end{tabular}                      &  \begin{tabular}[c]{@{}c@{}}38.84\\ \scriptsize{($\pm $0.13)}\end{tabular}            &  \begin{tabular}[c]{@{}c@{}}62.44\\ \scriptsize{($\pm $0.10)}\end{tabular}            & \begin{tabular}[c]{@{}c@{}}42.07\\ \scriptsize{($\pm $0.09)}\end{tabular}                     & \begin{tabular}[c]{@{}c@{}}\underline{65.12}\\ \scriptsize{($\pm $0.18)}\end{tabular}                     \\ 
			HUIL           &  \begin{tabular}[c]{@{}c@{}}40.25\\ \scriptsize{($\pm $0.33)}\end{tabular}            &  \begin{tabular}[c]{@{}c@{}}62.80\\ \scriptsize{($\pm $0.20)}\end{tabular}             & \begin{tabular}[c]{@{}c@{}}38.55\\ \scriptsize{($\pm $0.20)}\end{tabular}                     & \begin{tabular}[c]{@{}c@{}}59.90\\ \scriptsize{($\pm $0.20)}\end{tabular}                      &  \begin{tabular}[c]{@{}c@{}}38.65\\ \scriptsize{($\pm $0.17)}\end{tabular}            &  \begin{tabular}[c]{@{}c@{}}61.33\\ \scriptsize{($\pm $0.51)}\end{tabular}            & \begin{tabular}[c]{@{}c@{}}43.10\\ \scriptsize{($\pm $0.19)}\end{tabular}                     & \begin{tabular}[c]{@{}c@{}}63.52\\ \scriptsize{($\pm $0.42)}\end{tabular}                     \\ \hline
			\textbf{RCoCo} & \begin{tabular}[c]{@{}c@{}}\textbf{45.02}\\ \scriptsize{{($\pm $0.18)}}\end{tabular} & \begin{tabular}[c]{@{}c@{}}\textbf{64.85}\\ \scriptsize{{($\pm $0.30)}}\end{tabular} & \begin{tabular}[c]{@{}c@{}}\textbf{47.93}\\ \scriptsize{{($\pm $0.15)}}\end{tabular}  & \begin{tabular}[c]{@{}c@{}}\textbf{63.58}\\ \scriptsize{{($\pm $0.13)}}\end{tabular}  & \begin{tabular}[c]{@{}c@{}}\textbf{43.70}\\ \scriptsize{{($\pm $0.09)}}\end{tabular}  & \begin{tabular}[c]{@{}c@{}}\textbf{65.91}\\ \scriptsize{{($\pm $0.23)}}\end{tabular}  & \begin{tabular}[c]{@{}c@{}}\textbf{49.11}\\ \scriptsize{{($\pm $0.10)}}\end{tabular}  & \begin{tabular}[c]{@{}c@{}}\textbf{66.08}\\ \scriptsize{{($\pm $0.20)}}\end{tabular}  \\ \hline
	\end{tabular}}
\end{table*}

\begin{table*}[]
	\centering
	\caption{Inter-link prediction results on FB-TW$_A$, TW$_B$-FS, DBpedia$_\text{CH-EN}$, AMiner-DBLP datasets in terms of $MRR@K$ (\%), where standard derivations are given in brackets. The best results are in the \textbf{boldfaced}, and the second are \underline{underlined}.}
	\vspace{0.1in}
	\label{tab:t5}
	\resizebox{\linewidth}{!}{
		\begin{tabular}{c|cc|cc|cc|cc}
			\hline
			{\multirow{2}*{\diagbox[width=6.5em]{Method}{Dataset}}} & \multicolumn{2}{c|}{FB-TW$_A$} & \multicolumn{2}{c|}{TW$_B$-FS} & \multicolumn{2}{c|}{DBpedia$_\text{CH-EN}$} & \multicolumn{2}{c}{AMiner-DBLP} \\
			& $k$=10           & $k$=15          & $k$=10           & $k$=15          & $k$=10                 & $k$=15                 & $k$=10           & $k$=15           \\ \hline
			Moana          & \begin{tabular}[c]{@{}c@{}}31.54\\ \scriptsize{($\pm $0.19)}\end{tabular} & \begin{tabular}[c]{@{}c@{}}36.30\\ \scriptsize{($\pm $0.28)}\end{tabular} & \begin{tabular}[c]{@{}c@{}}32.12\\ \scriptsize{($\pm $0.15)}\end{tabular} & \begin{tabular}[c]{@{}c@{}}37.08\\ \scriptsize{($\pm $0.10)}\end{tabular} & \begin{tabular}[c]{@{}c@{}}36.32\\ \scriptsize{($\pm $0.11)}\end{tabular} & \begin{tabular}[c]{@{}c@{}}39.75\\ \scriptsize{($\pm $0.09)}\end{tabular} & \begin{tabular}[c]{@{}c@{}}33.58\\ \scriptsize{($\pm $0.21)}\end{tabular} & \begin{tabular}[c]{@{}c@{}}38.30\\ \scriptsize{($\pm $0.33)}\end{tabular} \\ 
			G-CREWE        & \begin{tabular}[c]{@{}c@{}}26.08\\ \scriptsize{($\pm $0.13)}\end{tabular} & \begin{tabular}[c]{@{}c@{}}28.75\\ \scriptsize{($\pm $0.31)}\end{tabular} & \begin{tabular}[c]{@{}c@{}}28.20\\ \scriptsize{($\pm $0.10)}\end{tabular}                                                            & \begin{tabular}[c]{@{}c@{}}31.22\\ \scriptsize{($\pm $0.22)}\end{tabular}                                                            & \begin{tabular}[c]{@{}c@{}}28.30\\ \scriptsize{($\pm $0.09)}\end{tabular} & \begin{tabular}[c]{@{}c@{}}34.36\\ \scriptsize{($\pm $0.23)}\end{tabular} & \begin{tabular}[c]{@{}c@{}}31.02\\ \scriptsize{($\pm $0.30)}\end{tabular}                                                            &\begin{tabular}[c]{@{}c@{}}35.16\\ \scriptsize{($\pm $0.12)}\end{tabular}                                                             \\ 
			SNNA           & \begin{tabular}[c]{@{}c@{}}34.10\\ \scriptsize{($\pm $0.17)}\end{tabular} & \begin{tabular}[c]{@{}c@{}}38.51\\ \scriptsize{($\pm $0.04)}\end{tabular} & \begin{tabular}[c]{@{}c@{}}33.57\\ \scriptsize{($\pm $0.12)}\end{tabular}                                                            & \begin{tabular}[c]{@{}c@{}}40.31\\ \scriptsize{($\pm $0.05)}\end{tabular}                                                            & \begin{tabular}[c]{@{}c@{}}40.21\\ \scriptsize{($\pm $0.29)}\end{tabular} & \begin{tabular}[c]{@{}c@{}}43.77\\ \scriptsize{($\pm $0.11)}\end{tabular} & \begin{tabular}[c]{@{}c@{}}32.92\\ \scriptsize{($\pm $0.03)}\end{tabular}                                                            &  \begin{tabular}[c]{@{}c@{}}38.21\\ \scriptsize{($\pm $0.20)}\end{tabular}                                                           \\ 
			Meta-NA        & \begin{tabular}[c]{@{}c@{}}28.99\\ \scriptsize{($\pm $0.28)}\end{tabular} & \begin{tabular}[c]{@{}c@{}}30.59\\ \scriptsize{($\pm $0.07)}\end{tabular} & \begin{tabular}[c]{@{}c@{}}30.52\\ \scriptsize{($\pm $0.20)}\end{tabular}                                                            & \begin{tabular}[c]{@{}c@{}}35.12\\ \scriptsize{($\pm $0.33)}\end{tabular}                                                            & \begin{tabular}[c]{@{}c@{}}33.84\\ \scriptsize{($\pm $0.15)}\end{tabular} & \begin{tabular}[c]{@{}c@{}}38.78\\ \scriptsize{($\pm $0.18)}\end{tabular} & \begin{tabular}[c]{@{}c@{}}29.22\\ \scriptsize{($\pm $0.50)}\end{tabular}                                                            & \begin{tabular}[c]{@{}c@{}}34.08\\ \scriptsize{($\pm $0.67)}\end{tabular}                                                            \\ 
			IONE-D         & \begin{tabular}[c]{@{}c@{}}34.01\\ \scriptsize{($\pm $0.22)}\end{tabular} & \begin{tabular}[c]{@{}c@{}}40.74\\ \scriptsize{($\pm $0.19)}\end{tabular} & \begin{tabular}[c]{@{}c@{}}34.75\\ \scriptsize{($\pm $0.52)}\end{tabular}                                                            & \begin{tabular}[c]{@{}c@{}}41.24\\ \scriptsize{($\pm $0.20)}\end{tabular}                                                            & \begin{tabular}[c]{@{}c@{}}42.67\\ \scriptsize{($\pm $0.26)}\end{tabular} & \begin{tabular}[c]{@{}c@{}}46.54\\ \scriptsize{($\pm $0.22)}\end{tabular} & \begin{tabular}[c]{@{}c@{}}35.15\\ \scriptsize{($\pm $0.13)}\end{tabular}                                         &  \begin{tabular}[c]{@{}c@{}}42.33\\ \scriptsize{($\pm $0.35)}\end{tabular}                                                             \\ 
			NeXtAlign      & \begin{tabular}[c]{@{}c@{}}$\underline{36.73}$\\ \scriptsize{{($\pm $0.28)}}\end{tabular} & \begin{tabular}[c]{@{}c@{}}\underline{43.26}\\ \scriptsize{($\pm $0.17)}\end{tabular} & \begin{tabular}[c]{@{}c@{}}37.06\\ \scriptsize{($\pm $0.33)}\end{tabular}                                                            & \begin{tabular}[c]{@{}c@{}}\underline{44.50}\\ \scriptsize{($\pm $0.15)}\end{tabular}                                                              & \begin{tabular}[c]{@{}c@{}}\underline{45.50}\\ \scriptsize{($\pm $0.19)}\end{tabular} & \begin{tabular}[c]{@{}c@{}}\underline{48.67}\\ \scriptsize{($\pm $0.21)}\end{tabular} & \begin{tabular}[c]{@{}c@{}}\underline{39.10}\\ \scriptsize{($\pm $0.33)}\end{tabular}                                                              & \begin{tabular}[c]{@{}c@{}}45.22\\ \scriptsize{($\pm $0.10)}\end{tabular}                                                              \\ 
			NAME           & \begin{tabular}[c]{@{}c@{}}35.59\\ \scriptsize{($\pm $0.18)}\end{tabular} & \begin{tabular}[c]{@{}c@{}}42.91\\ \scriptsize{($\pm $0.16)}\end{tabular} & \begin{tabular}[c]{@{}c@{}}\underline{37.15}\\ \scriptsize{($\pm $0.12)}\end{tabular}                                                              & \begin{tabular}[c]{@{}c@{}}43.86\\ \scriptsize{($\pm $0.10)}\end{tabular}                                                              & \begin{tabular}[c]{@{}c@{}}44.61\\ \scriptsize{($\pm $0.18)}\end{tabular} & \begin{tabular}[c]{@{}c@{}}47.23\\ \scriptsize{($\pm $0.17)}\end{tabular} &  \begin{tabular}[c]{@{}c@{}}38.51\\ \scriptsize{($\pm $0.20)}\end{tabular}                                                             & \begin{tabular}[c]{@{}c@{}}\underline{45.60}\\ \scriptsize{($\pm $0.15)}\end{tabular}                                                              \\ 
			HUIL           &  \begin{tabular}[c]{@{}c@{}}35.80\\ \scriptsize{($\pm $0.13)}\end{tabular}            &  \begin{tabular}[c]{@{}c@{}}42.56\\ \scriptsize{($\pm $0.12)}\end{tabular}             & \begin{tabular}[c]{@{}c@{}}36.33\\ \scriptsize{($\pm $0.30)}\end{tabular}                        & \begin{tabular}[c]{@{}c@{}}42.16\\ \scriptsize{($\pm $0.08)}\end{tabular}                         &  \begin{tabular}[c]{@{}c@{}}39.65\\ \scriptsize{($\pm $0.10)}\end{tabular}            &  \begin{tabular}[c]{@{}c@{}}44.20\\ \scriptsize{($\pm $0.33)}\end{tabular}            & \begin{tabular}[c]{@{}c@{}}37.79\\ \scriptsize{($\pm $0.06)}\end{tabular}                        & \begin{tabular}[c]{@{}c@{}}43.15\\ \scriptsize{($\pm $0.30)}\end{tabular}                        \\ \hline
			\textbf{RCoCo} & \begin{tabular}[c]{@{}c@{}}\textbf{38.72}\\ \scriptsize{{($\pm $0.20)}}\end{tabular}                                                           &  \begin{tabular}[c]{@{}c@{}}\textbf{46.02}\\ \scriptsize{{($\pm $0.33)}}\end{tabular}                                                          & \begin{tabular}[c]{@{}c@{}}\textbf{39.30}\\ \scriptsize{{($\pm $0.25)}}\end{tabular}                                                           & \begin{tabular}[c]{@{}c@{}}\textbf{46.29}\\ \scriptsize{{($\pm $0.17)}}\end{tabular}                                                           & \begin{tabular}[c]{@{}c@{}}\textbf{46.02}\\ \scriptsize{{($\pm $0.20)}}\end{tabular}                                                           & \begin{tabular}[c]{@{}c@{}}\textbf{49.27}\\ \scriptsize{{($\pm $0.15)}}\end{tabular}                                                           & \begin{tabular}[c]{@{}c@{}}\textbf{40.56}\\ \scriptsize{{($\pm $0.21)}}\end{tabular}                                                           & \begin{tabular}[c]{@{}c@{}}\textbf{47.20}\\ \scriptsize{{($\pm $0.11)}}\end{tabular}                                                           \\ \hline
	\end{tabular}}
\end{table*}

\subsection{Ablation Study} 

In this section, we examine the effectiveness of the two key design in the proposed RCoCo, the curvature-awareness of $\kappa-$GAT and the community-based graph augmentation.

To this end, we design two groups of variant models. The first group is on the curvature of the manifold. Specifically, without the curvature estimation in $\kappa-$GAT, we use the predefined curvature instead, and instantiate a $1-$GAT of the curvature $\kappa=1$ for hyperspherical space, a $(-1)-$GAT of the curvature $\kappa=-1$ for hyperbolic space, and $0-$GAT for the zero-curvature Euclidean space. Here, we employ the standard curvature of each type of manifold as a representative. As social networks are usually connected by sharing similarities, the curvature of both layers is set as the predefined value. The variants are denoted by the curvature ($1$, $0$ and $-1$).
The second group is on the community-based graph augmentation in which we disable node-to-supernode contrast ($\mathcal L_{n2s}$) in the intra-contrastive learning and supernode discovery loss $\mathcal Q$, denoted as $-\mathcal L_{n2s}$ variant.
That is, the model is trained by the node-to-node contrastive and inter-contrastive loss only.
Accordingly, we combine different curvatures and loss formulations, and obtain \textbf{6 variants} in total.

Without loss of generality, we conduct ablation study in both intra-link prediction and inter-link prediction.
Table \ref{tab:12} and Table \ref{tab:11} record the empirical results of intra-link prediction on all the 8 datasets (in terms of AUC) and inter-link prediction on the 4 network pairs (in terms of MRR), respectively.
As shown in the tables, we find that: 
\begin{itemize}
\item \emph{The $-\mathcal L_{n2s}$ variant has inferior performance regardless of curvature, verifying the effectiveness of community-based graph augmentation. }The community reveals the organization and structural pattern of the social network, and node-to-community contrast tends to generate more informative user representations. Also, it suggests that the community can be explored in graph augmentation for general purpose.
\item \emph{$\kappa-$GAT consistently achieves superior performance to variants of predefined curvatures}, and the variants obtain their best results on different datasets as shown in Table \ref{tab:12}. For instance, the variant with $1-$GAT achieves competitive AUC to RCoCo on DBpedia$_{\text{CH}}$ as well as DBpedia$_{\text{EN}}$, while presents significant AUC loss on the other datasets. Recall the empirical investigation on the geometry of the datasets in Table \ref{tab:example}. It shows that a learning model tend to have better expressiveness when its geometric (inductive) bias of curvature matches the inherent structure of the datasets. In other words, it demonstrates the necessity of designing curvature estimator in RCoCo. In addition, RCoCo achieves different performance gains on the same network pair, e.g., DBpedia$_{\text{CH}}$ and DBpedia$_{\text{EN}}$. That is, curvatures of different layers in the same multiplex network can still be different to some extent, verifying the motivation of our study. This is also the reason why we utilize independent curvature estimator for each layer of social network in RCoCo.
\end{itemize}

\begin{table*}[]
	\centering
	\caption{Ablation study of RCoCo for link prediction results on FB, TW$_A$, TW$_B$, FS, DBpedia$_{\text{CH}}$, DBpedia$_{\text{EN}}$, AMiner and DBLP datasets in terms of AUC (\%), where standard derivations are given in brackets. The best results are in the \textbf{boldfaced}, and the second are \underline{underlined}.}
	\vspace{0.1in}
	\label{tab:12}
	\resizebox{\linewidth}{!}{
	\begin{tabular}{cc|c|c|c|c|c|c|c|c}
		\hline
		\multicolumn{2}{c|}{\diagbox[width=6.5em]{Variant}{Dataset}}                                          & FB                                                           & TW$_A$                                                       & TW$_B$                                                       & FS                                                           & DB$_\text{CH}$                                               & DB$_\text{EN}$                                               & AMiner                                                       & DBLP                                                         \\ \hline

		\multicolumn{1}{c|}{\multirow{2}{*}{\rotatebox{90}{$\kappa-$GAT}}} & $Origin$  & \begin{tabular}[c]{@{}c@{}}\textbf{92.24}\\ ($\pm ${0.11})\end{tabular} & \begin{tabular}[c]{@{}c@{}}\textbf{92.63}\\ ($\pm ${0.20})\end{tabular} & \begin{tabular}[c]{@{}c@{}}\textbf{92.16}\\ ($\pm ${0.07})\end{tabular} & \begin{tabular}[c]{@{}c@{}}\textbf{91.59}\\ ($\pm ${0.25})\end{tabular} & \begin{tabular}[c]{@{}c@{}}\textbf{90.76}\\ ($\pm ${0.22})\end{tabular} & \begin{tabular}[c]{@{}c@{}}\textbf{90.69}\\ ($\pm ${0.20})\end{tabular} & \begin{tabular}[c]{@{}c@{}}\textbf{93.25}\\ ($\pm ${0.03})\end{tabular} & \begin{tabular}[c]{@{}c@{}}\textbf{93.97}\\ ($\pm ${0.20})\end{tabular}  \\
		\multicolumn{1}{c|}{}                          & $-\mathcal{L}_{n2s}$ & \begin{tabular}[c]{@{}c@{}}91.79\\ ($\pm $0.19)\end{tabular} & \begin{tabular}[c]{@{}c@{}}91.96\\ ($\pm $0.13)\end{tabular} & \begin{tabular}[c]{@{}c@{}}91.55\\ ($\pm $0.10)\end{tabular} & \begin{tabular}[c]{@{}c@{}}90.82\\ ($\pm $0.22)\end{tabular} & \begin{tabular}[c]{@{}c@{}}89.37\\ ($\pm $0.23)\end{tabular} & \begin{tabular}[c]{@{}c@{}}\underline{90.25}\\ ($\pm $0.16)\end{tabular} & \begin{tabular}[c]{@{}c@{}}92.36\\ ($\pm $0.21)\end{tabular} & \begin{tabular}[c]{@{}c@{}}92.70\\ ($\pm $0.11)\end{tabular} \\ \hline

		\multicolumn{1}{c|}{\multirow{2}{*}{\rotatebox{90}{1-GAT}}}    & $Origin$  & \begin{tabular}[c]{@{}c@{}}89.75\\ ($\pm $0.20)\end{tabular}  & \begin{tabular}[c]{@{}c@{}}90.51\\ ($\pm $0.18)\end{tabular} & \begin{tabular}[c]{@{}c@{}}89.67\\ ($\pm $0.21)\end{tabular} & \begin{tabular}[c]{@{}c@{}}90.52\\ ($\pm $0.19)\end{tabular} & \begin{tabular}[c]{@{}c@{}}\underline{90.25}\\ ($\pm $0.31)\end{tabular} & \begin{tabular}[c]{@{}c@{}}{90.16}\\ ($\pm $1.02)\end{tabular} & \begin{tabular}[c]{@{}c@{}}90.12\\ ($\pm $0.28)\end{tabular} & \begin{tabular}[c]{@{}c@{}}89.63\\ ($\pm $0.14)\end{tabular} \\
		\multicolumn{1}{c|}{}                          & $-\mathcal{L}_{n2s}$ & \begin{tabular}[c]{@{}c@{}}87.56\\ ($\pm $0.15)\end{tabular} & \begin{tabular}[c]{@{}c@{}}88.79\\ ($\pm $0.11)\end{tabular} & \begin{tabular}[c]{@{}c@{}}87.33\\ ($\pm $0.16)\end{tabular} & \begin{tabular}[c]{@{}c@{}}89.07\\ ($\pm $0.10)\end{tabular} & \begin{tabular}[c]{@{}c@{}}89.02\\ ($\pm $0.24)\end{tabular} & \begin{tabular}[c]{@{}c@{}}88.73\\ ($\pm $0.08)\end{tabular} & \begin{tabular}[c]{@{}c@{}}88.59\\ ($\pm $0.15)\end{tabular} & \begin{tabular}[c]{@{}c@{}}88.51\\ ($\pm $0.21)\end{tabular} \\ \hline

		\multicolumn{1}{c|}{\multirow{2}{*}{\rotatebox{90}{0-GAT}}}    & $Origin$ & \begin{tabular}[c]{@{}c@{}}89.03\\ ($\pm $0.33)\end{tabular} & \begin{tabular}[c]{@{}c@{}}88.62\\ ($\pm $0.10)\end{tabular} & \begin{tabular}[c]{@{}c@{}}89.25\\ ($\pm $0.18)\end{tabular} & \begin{tabular}[c]{@{}c@{}}88.73\\ ($\pm $0.15)\end{tabular} & \begin{tabular}[c]{@{}c@{}}87.67\\ ($\pm $0.11)\end{tabular} & \begin{tabular}[c]{@{}c@{}}88.90\\ ($\pm $0.56)\end{tabular} & \begin{tabular}[c]{@{}c@{}}92.12\\ ($\pm $0.11)\end{tabular} & \begin{tabular}[c]{@{}c@{}}92.27\\ ($\pm $0.29)\end{tabular} \\
		\multicolumn{1}{c|}{}                          & $-\mathcal{L}_{n2s}$ & \begin{tabular}[c]{@{}c@{}}88.12\\ ($\pm $0.10)\end{tabular} & \begin{tabular}[c]{@{}c@{}}87.90\\ ($\pm $0.25)\end{tabular}  & \begin{tabular}[c]{@{}c@{}}88.18\\ ($\pm $0.11)\end{tabular} & \begin{tabular}[c]{@{}c@{}}88.12\\ ($\pm $0.22)\end{tabular} & \begin{tabular}[c]{@{}c@{}}85.91\\ ($\pm $0.16)\end{tabular} & \begin{tabular}[c]{@{}c@{}}87.62\\ ($\pm $0.21)\end{tabular} & \begin{tabular}[c]{@{}c@{}}91.77\\ ($\pm $0.10)\end{tabular} & \begin{tabular}[c]{@{}c@{}}91.89\\ ($\pm $0.36)\end{tabular} \\ \hline
		
		\multicolumn{1}{c|}{\multirow{2}{*}{\rotatebox{90}{(-1)-GAT}}} & $Origin$  & \begin{tabular}[c]{@{}c@{}}$\underline{91.85}$\\ ($\pm $0.08)\end{tabular} & \begin{tabular}[c]{@{}c@{}}$\underline{92.59}$\\ ($\pm $0.71)\end{tabular} & \begin{tabular}[c]{@{}c@{}}\underline{91.70}\\ ($\pm $0.24)\end{tabular} & \begin{tabular}[c]{@{}c@{}}\underline{91.36}\\ ($\pm $0.20)\end{tabular} & \begin{tabular}[c]{@{}c@{}}88.92\\ ($\pm $0.10)\end{tabular} & \begin{tabular}[c]{@{}c@{}}88.53\\ ($\pm $0.10)\end{tabular} & \begin{tabular}[c]{@{}c@{}}\underline{92.61}\\ ($\pm $0.25)\end{tabular} & \begin{tabular}[c]{@{}c@{}}\underline{93.01}\\ ($\pm $0.06)\end{tabular} \\
		\multicolumn{1}{c|}{}                          & $-\mathcal{L}_{n2s}$ & \begin{tabular}[c]{@{}c@{}}89.93\\ ($\pm $0.15)\end{tabular} & \begin{tabular}[c]{@{}c@{}}91.27\\ ($\pm $0.10)\end{tabular} & \begin{tabular}[c]{@{}c@{}}90.16\\ ($\pm $0.11)\end{tabular} & \begin{tabular}[c]{@{}c@{}}89.96\\ ($\pm $0.13)\end{tabular} & \begin{tabular}[c]{@{}c@{}}88.06\\ ($\pm $0.11)\end{tabular} & \begin{tabular}[c]{@{}c@{}}86.91\\ ($\pm $0.37)\end{tabular} & \begin{tabular}[c]{@{}c@{}}91.82\\ ($\pm $0.18)\end{tabular} & \begin{tabular}[c]{@{}c@{}}91.96\\ ($\pm $0.12)\end{tabular} \\ \hline
		\end{tabular}}
\end{table*}

\begin{table*}[]
	\centering		
	\caption{Ablation study of RCoCo for user alignment results on FB-TW$_A$, TW$_B$-FS, DBpedia$_\text{CH-EN}$, AMiner-DBLP datasets in terms of MRR-k (\%), where standard derivations are given in brackets. The best results are in the \textbf{boldfaced}, and the second are \underline{underlined}.}
	\vspace{0.1in}
	\resizebox{\linewidth}{!}
	{
		\label{tab:11}
		\begin{tabular}{cc|cc|cc|cc|cc}
			\hline
			\multicolumn{2}{c|}{\multirow{2}*{\diagbox[width=6.5em]{Variant}{Dataset}}} & \multicolumn{2}{c|}{FB-TW$_A$}                                                                                                                & \multicolumn{2}{c|}{TW$_B$-FS}                                                                                                                & \multicolumn{2}{c|}{DBpedia$_\text{CH-EN}$}                                                                                                   & \multicolumn{2}{c}{AMiner-DBLP}                                                                                                               \\ 
			\multicolumn{2}{c|}{}                                                               & $k$=10                                                                  & $k$=15                                                                  & $k$=10                                                                  & $k$=15                                                                  & $k$=10                                                                  & $k$=15                                                                  & $k$=10                                                                  & $k$=15                                                                  \\ \hline

			\multicolumn{1}{c|}{\multirow{2}{*}{\rotatebox{90}{$\kappa-$GAT}}} & $Origin$  & {\begin{tabular}[c]{@{}c@{}}\textbf{38.72}\\ ($\pm $0.20)\end{tabular}} & {\begin{tabular}[c]{@{}c@{}}\textbf{46.02}\\ ($\pm $0.33)\end{tabular}} & {\begin{tabular}[c]{@{}c@{}}\textbf{39.30}\\ ($\pm $0.25)\end{tabular}} & {\begin{tabular}[c]{@{}c@{}}\textbf{46.29}\\ ($\pm $0.17)\end{tabular}} & {\begin{tabular}[c]{@{}c@{}}\textbf{35.92}\\ ($\pm $0.20)\end{tabular}} & {\begin{tabular}[c]{@{}c@{}}\textbf{49.27}\\ ($\pm $0.15)\end{tabular}} & {\begin{tabular}[c]{@{}c@{}}\textbf{40.56}\\ ($\pm $0.21)\end{tabular}} & {\begin{tabular}[c]{@{}c@{}}\textbf{47.20}\\ ($\pm $0.11)\end{tabular}}          \\
			\multicolumn{1}{c|}{}                          & $-\mathcal{L}_{n2s}$ & \begin{tabular}[c]{@{}c@{}}38.39\\ ($\pm $0.37)\end{tabular}          & \begin{tabular}[c]{@{}c@{}}45.16\\ ($\pm $0.26)\end{tabular}          & \begin{tabular}[c]{@{}c@{}}38.11\\ ($\pm $0.16)\end{tabular}          & \begin{tabular}[c]{@{}c@{}}44.93\\ ($\pm $0.29)\end{tabular}          & \begin{tabular}[c]{@{}c@{}}34.15\\ ($\pm $0.13)\end{tabular}          & \begin{tabular}[c]{@{}c@{}}48.06\\ ($\pm $0.33)\end{tabular}          & \begin{tabular}[c]{@{}c@{}}39.18\\ ($\pm $0.08)\end{tabular}          & \begin{tabular}[c]{@{}c@{}}46.12\\ ($\pm $0.26)\end{tabular}          \\\hline

			\multicolumn{1}{c|}{\multirow{2}{*}{\rotatebox{90}{1-GAT}}}    & $Origin$ & \begin{tabular}[c]{@{}c@{}}37.26\\ ($\pm $0.23)\end{tabular}          & \begin{tabular}[c]{@{}c@{}}44.12\\ ($\pm $0.06)\end{tabular}          & \begin{tabular}[c]{@{}c@{}}37.11\\ ($\pm $0.13)\end{tabular}          & \begin{tabular}[c]{@{}c@{}}44.71\\ ($\pm $0.22)\end{tabular}          & \begin{tabular}[c]{@{}c@{}}\underline{35.60}\\ ($\pm $0.10)\end{tabular}          & \begin{tabular}[c]{@{}c@{}}\underline{48.86}\\ ($\pm $0.17)\end{tabular}          & \begin{tabular}[c]{@{}c@{}}38.71\\ ($\pm $0.06)\end{tabular}          & \begin{tabular}[c]{@{}c@{}}45.05\\ ($\pm $0.21)\end{tabular}          \\
			\multicolumn{1}{c|}{}                          & $-\mathcal{L}_{n2s}$ & \begin{tabular}[c]{@{}c@{}}35.05\\ ($\pm $0.12)\end{tabular}          & \begin{tabular}[c]{@{}c@{}}43.36\\ ($\pm $1.05)\end{tabular}          & \begin{tabular}[c]{@{}c@{}}36.29\\ ($\pm $0.26)\end{tabular}          & \begin{tabular}[c]{@{}c@{}}43.81\\ ($\pm $0.21)\end{tabular}          & \begin{tabular}[c]{@{}c@{}}33.73\\ ($\pm $0.11)\end{tabular}          & \begin{tabular}[c]{@{}c@{}}47.19\\ ($\pm $0.13)\end{tabular}          & \begin{tabular}[c]{@{}c@{}}37.24\\ ($\pm $0.15)\end{tabular}          & \begin{tabular}[c]{@{}c@{}}44.10\\ ($\pm $0.29)\end{tabular}          \\ \hline
			\multicolumn{1}{c|}{\multirow{2}{*}{\rotatebox{90}{0-GAT}}}    & $Origin$  & \begin{tabular}[c]{@{}c@{}}36.91\\ ($\pm $0.30)\end{tabular}          & \begin{tabular}[c]{@{}c@{}}43.08\\ ($\pm $0.21)\end{tabular}          & \begin{tabular}[c]{@{}c@{}}36.67\\ ($\pm $0.12)\end{tabular}          & \begin{tabular}[c]{@{}c@{}}44.16\\ ($\pm $0.11)\end{tabular}          & \begin{tabular}[c]{@{}c@{}}32.45\\ ($\pm $0.17)\end{tabular}          & \begin{tabular}[c]{@{}c@{}}47.07\\ ($\pm $0.12)\end{tabular}          & \begin{tabular}[c]{@{}c@{}}39.33\\ ($\pm $0.21)\end{tabular}          & \begin{tabular}[c]{@{}c@{}}45.81\\ ($\pm $0.15)\end{tabular}          \\
			\multicolumn{1}{c|}{}                          & $-\mathcal{L}_{n2s}$ & \begin{tabular}[c]{@{}c@{}}35.24\\ ($\pm $0.16)\end{tabular}          & \begin{tabular}[c]{@{}c@{}}41.80\\ ($\pm $0.19)\end{tabular}          & \begin{tabular}[c]{@{}c@{}}35.25\\ ($\pm $0.51)\end{tabular}          & \begin{tabular}[c]{@{}c@{}}42.57\\ ($\pm $0.14)\end{tabular}          & \begin{tabular}[c]{@{}c@{}}32.02\\ ($\pm $0.26)\end{tabular}          & \begin{tabular}[c]{@{}c@{}}45.93\\ ($\pm $0.18)\end{tabular}          & \begin{tabular}[c]{@{}c@{}}38.58\\ ($\pm $0.15)\end{tabular}          & \begin{tabular}[c]{@{}c@{}}43.67\\ ($\pm $0.17)\end{tabular}          \\ \hline
			
			\multicolumn{1}{c|}{\multirow{2}{*}{\rotatebox{90}{(-1)-GAT}}} & $Origin$  & \begin{tabular}[c]{@{}c@{}}\underline{38.50}\\ ($\pm $0.28)\end{tabular}          & \begin{tabular}[c]{@{}c@{}}\underline{45.59}\\ ($\pm $0.31)\end{tabular}          & \begin{tabular}[c]{@{}c@{}}\underline{38.32}\\ ($\pm $0.18)\end{tabular}          & \begin{tabular}[c]{@{}c@{}}\underline{45.24}\\ ($\pm $0.09)\end{tabular}          & \begin{tabular}[c]{@{}c@{}}33.96\\ ($\pm $0.56)\end{tabular}          & \begin{tabular}[c]{@{}c@{}}47.91\\ ($\pm $0.21)\end{tabular}          & \begin{tabular}[c]{@{}c@{}}\underline{40.12}\\ ($\pm $0.10)\end{tabular}          & \begin{tabular}[c]{@{}c@{}}\underline{46.85}\\ ($\pm $0.67)\end{tabular}          \\
			\multicolumn{1}{c|}{}                          & $-\mathcal{L}_{n2s}$ & \begin{tabular}[c]{@{}c@{}}37.03\\ ($\pm $0.55)\end{tabular}          & \begin{tabular}[c]{@{}c@{}}43.23\\ ($\pm $0.23)\end{tabular}          & \begin{tabular}[c]{@{}c@{}}36.93\\ ($\pm $0.20)\end{tabular}          & \begin{tabular}[c]{@{}c@{}}43.95\\ ($\pm $0.37)\end{tabular}          & \begin{tabular}[c]{@{}c@{}}32.82\\ ($\pm $0.33)\end{tabular}          & \begin{tabular}[c]{@{}c@{}}46.25\\ ($\pm $0.17)\end{tabular}          & \begin{tabular}[c]{@{}c@{}}38.93\\ ($\pm $0.22)\end{tabular}          & \begin{tabular}[c]{@{}c@{}}45.17\\ ($\pm $0.19)\end{tabular}          \\ \hline
			\end{tabular}}
\end{table*}

\subsection{Sensitivity on Network Overlap}
The network overlap $(\eta)$ is a key hyper-parameter to user alignment, and we study the parameter sensitvity of $\eta$ in this section. 

Network Overlap is given as $\eta=\frac{2 \times (\text{\# of anchor users})}{\text{\# of node in }\mathcal{G}_{1} + \text{\# of node in }\mathcal{G}_{2}}$, and  we use 20\% overlap as defult.
We vary the network overlap of RCoCo in \{20\%,15\%,10\%,5\%\}. On the one hand, we do user alignment prediction via RCoCo on the FB-TW$_A$, TW$_B$-FS, DBpedia$_\text{CH-EN}$, AMiner-DBLP datasets, and report the empirical results of $Hit@K(\%)$ and $MRR@K(\%)$ in Table \ref{tab:t01} and \ref{tab:t02}, respectively. As the network overlap increases, the performance of RCoCo shows a significant improvement. It suggests that 1) A higher network overlap indicates a greater number of nodes pairs between the networks, which is helpful to establish a corresponding relationship between networks. 
2) When network overlap is high, nodes pairs and similarity information enable the restoration and alignment of networks even in the noise or missing nodes.

On the other hand, we compare the RCoCo method with baselines on the TW$_B$-FS dataset and summarize the prediction results as shown in Table \ref{tab:t03}, where $k$ is set to 10. Compared to baselines, RCoCo performs well in alignment prediction for different network overlaps. This indicates that RCoCo has higher accuracy and performance when dealing with user alignment problems. The main reason is that 1) RCoCo utilizes graph augmentations in constant curvature space to reduce the dependence in anchor nodes. 2) RCoCo learns node similarity informations between networks through reasonable neighbor selection and feature learning. Our method is able to capture the similarities and correspondences accurately between networks, improving prediction accuracy.

\begin{table*}[]
	\centering
	\caption{Network Overlap study of RCoCo for user alignment results on FB-TW$_A$, TW$_B$-FS, DBpedia$_\text{CH-EN}$, AMiner-DBLP datasets in terms of $Hit@K$ (\%), where standard derivations are given in brackets. The best results are in the \textbf{boldfaced}, and the second are \underline{underlined}.}
	\vspace{0.1in}
	\label{tab:t01}
	\resizebox{\linewidth}{!}{
		\begin{tabular}{c|cc|cc|cc|cc}
			\hline
			{\multirow{2}*{\diagbox[width=5.8em]{Overlap}{Dataset}}} & \multicolumn{2}{c|}{FB-TW$_A$} & \multicolumn{2}{c|}{TW$_B$-FS} & \multicolumn{2}{c|}{DBpedia$_\text{CH-EN}$} & \multicolumn{2}{c}{AMiner-DBLP} \\
			& $k$=10           & $k$=15          & $k$=10           & $k$=15          & $k$=10                 & $k$=15                 & $k$=10           & $k$=15           \\ \hline

			$\eta =$20\%          & \begin{tabular}[c]{@{}c@{}}\textbf{45.02}\\ \scriptsize{($\pm $0.18)}\end{tabular}           & \begin{tabular}[c]{@{}c@{}}\textbf{64.85}\\ \scriptsize{($\pm $0.30)}\end{tabular}            &\begin{tabular}[c]{@{}c@{}}\textbf{47.93}\\ \scriptsize{($\pm $0.15)}\end{tabular}                      &\begin{tabular}[c]{@{}c@{}}\textbf{63.58}\\ \scriptsize{($\pm $0.13)}\end{tabular}                       & \begin{tabular}[c]{@{}c@{}}\textbf{43.70}\\ \scriptsize{($\pm $0.09)}\end{tabular}           & \begin{tabular}[c]{@{}c@{}}\textbf{65.91}\\ \scriptsize{($\pm $0.23)}\end{tabular}            &\begin{tabular}[c]{@{}c@{}}\textbf{49.11}\\ \scriptsize{($\pm $0.10)}\end{tabular}                      &\begin{tabular}[c]{@{}c@{}}\textbf{66.08}\\ \scriptsize{($\pm $0.20)}\end{tabular}                       \\

			$\eta =$15\%        & \begin{tabular}[c]{@{}c@{}}\underline{41.27}\\ \scriptsize{($\pm $0.23)}\end{tabular}            & \begin{tabular}[c]{@{}c@{}}\underline{57.33}\\ \scriptsize{($\pm $0.28)}\end{tabular}             & \begin{tabular}[c]{@{}c@{}}\underline{43.35}\\ \scriptsize{($\pm $0.39)}\end{tabular}                     & \begin{tabular}[c]{@{}c@{}}\underline{55.91}\\ \scriptsize{($\pm $0.27)}\end{tabular}                      & \begin{tabular}[c]{@{}c@{}}\underline{39.25}\\ \scriptsize{($\pm $0.31)}\end{tabular}            & \begin{tabular}[c]{@{}c@{}}\underline{60.12}\\ \scriptsize{($\pm $0.45)}\end{tabular}             & \begin{tabular}[c]{@{}c@{}}\underline{41.52}\\ \scriptsize{($\pm $0.19)}\end{tabular}                     & \begin{tabular}[c]{@{}c@{}}\underline{61.27}\\ \scriptsize{($\pm $0.32)}\end{tabular}                     \\ 
			$\eta =$10\%           & \begin{tabular}[c]{@{}c@{}}32.52\\ \scriptsize{($\pm $0.57)}\end{tabular}            &  \begin{tabular}[c]{@{}c@{}}49.02\\ \scriptsize{($\pm $0.64)}\end{tabular}            & \begin{tabular}[c]{@{}c@{}}37.20\\ \scriptsize{($\pm $1.03)}\end{tabular}                     & \begin{tabular}[c]{@{}c@{}}48.20\\ \scriptsize{($\pm $0.72)}\end{tabular}                      &  \begin{tabular}[c]{@{}c@{}}31.02\\ \scriptsize{($\pm $1.01)}\end{tabular}            &  \begin{tabular}[c]{@{}c@{}}47.81\\ \scriptsize{($\pm $1.12)}\end{tabular}             & \begin{tabular}[c]{@{}c@{}}33.29\\ \scriptsize{($\pm $0.93)}\end{tabular}                     & \begin{tabular}[c]{@{}c@{}}45.79\\ \scriptsize{($\pm $0.61)}\end{tabular}                     \\

			$\eta =$5\%      &  \begin{tabular}[c]{@{}c@{}}29.13\\ \scriptsize{($\pm $0.11)}\end{tabular}            &  \begin{tabular}[c]{@{}c@{}}32.30\\ \scriptsize{($\pm $0.18)}\end{tabular}                                & \begin{tabular}[c]{@{}c@{}}27.08\\ \scriptsize{($\pm $0.03)}\end{tabular}                      &  \begin{tabular}[c]{@{}c@{}}33.16\\ \scriptsize{($\pm $0.10)}\end{tabular}            &  \begin{tabular}[c]{@{}c@{}}28.55\\ \scriptsize{($\pm $0.12)}\end{tabular}             & \begin{tabular}[c]{@{}c@{}}33.12\\ \scriptsize{($\pm $0.07)}\end{tabular}                     & \begin{tabular}[c]{@{}c@{}}29.60\\ \scriptsize{($\pm $0.17)}\end{tabular}                      & \begin{tabular}[c]{@{}c@{}}36.18\\ \scriptsize{($\pm $0.11)}\end{tabular} \\ 
			                     
		   \hline
	\end{tabular}}
\end{table*}

\begin{table*}[]
	\centering
	\caption{Network Overlap study of RCoCo for user alignment results on FB-TW$_A$, TW$_B$-FS, DBpedia$_\text{CH-EN}$, AMiner-DBLP datasets in terms of $MRR@K$ (\%), where standard derivations are given in brackets. The best results are in the \textbf{boldfaced}, and the second are \underline{underlined}.}
	\vspace{0.1in}
	\label{tab:t02}
	\resizebox{\linewidth}{!}{
		\begin{tabular}{c|cc|cc|cc|cc}
			\hline
			{\multirow{2}*{\diagbox[width=5.8em]{Overlap}{Dataset}}} & \multicolumn{2}{c|}{FB-TW$_A$} & \multicolumn{2}{c|}{TW$_B$-FS} & \multicolumn{2}{c|}{DBpedia$_\text{CH-EN}$} & \multicolumn{2}{c}{AMiner-DBLP} \\
			& $k$=10           & $k$=15          & $k$=10           & $k$=15          & $k$=10                 & $k$=15                 & $k$=10           & $k$=15           \\ \hline

			$\eta =$20\%          & \begin{tabular}[c]{@{}c@{}}\textbf{38.72}\\ \scriptsize{($\pm $0.20)}\end{tabular}           & \begin{tabular}[c]{@{}c@{}}\textbf{46.02}\\ \scriptsize{($\pm $0.33)}\end{tabular}            &\begin{tabular}[c]{@{}c@{}}\textbf{39.30}\\ \scriptsize{($\pm $0.25)}\end{tabular}                      &\begin{tabular}[c]{@{}c@{}}\textbf{46.29}\\ \scriptsize{($\pm $0.17)}\end{tabular}                       & \begin{tabular}[c]{@{}c@{}}\textbf{46.02}\\ \scriptsize{($\pm $0.20)}\end{tabular}           & \begin{tabular}[c]{@{}c@{}}\textbf{49.27}\\ \scriptsize{($\pm $0.15)}\end{tabular}            &\begin{tabular}[c]{@{}c@{}}\textbf{40.56}\\ \scriptsize{($\pm $0.21)}\end{tabular}                      &\begin{tabular}[c]{@{}c@{}}\textbf{47.20}\\ \scriptsize{($\pm $0.11)}\end{tabular}                       \\

			$\eta =$15\%        & \begin{tabular}[c]{@{}c@{}}\underline{29.17}\\ \scriptsize{($\pm $0.36)}\end{tabular}            & \begin{tabular}[c]{@{}c@{}}\underline{41.35}\\ \scriptsize{($\pm $0.30)}\end{tabular}             & \begin{tabular}[c]{@{}c@{}}\underline{31.82}\\ \scriptsize{($\pm $0.15)}\end{tabular}                     & \begin{tabular}[c]{@{}c@{}}\underline{39.26}\\ \scriptsize{($\pm $0.18)}\end{tabular}                      & \begin{tabular}[c]{@{}c@{}}\underline{31.22}\\ \scriptsize{($\pm $0.31)}\end{tabular}            & \begin{tabular}[c]{@{}c@{}}\underline{41.93}\\ \scriptsize{($\pm $0.22)}\end{tabular}             & \begin{tabular}[c]{@{}c@{}}\underline{32.06}\\ \scriptsize{($\pm $0.19)}\end{tabular}                     & \begin{tabular}[c]{@{}c@{}}\underline{42.81}\\ \scriptsize{($\pm $0.30)}\end{tabular}                     \\ 
			
			$\eta =$10\%           & \begin{tabular}[c]{@{}c@{}}23.68\\ \scriptsize{($\pm $0.33)}\end{tabular}            &  \begin{tabular}[c]{@{}c@{}}29.03\\ \scriptsize{($\pm $0.57)}\end{tabular}            & \begin{tabular}[c]{@{}c@{}}23.05\\ \scriptsize{($\pm $0.35)}\end{tabular}                     & \begin{tabular}[c]{@{}c@{}}28.67\\ \scriptsize{($\pm $0.26)}\end{tabular}                      &  \begin{tabular}[c]{@{}c@{}}25.36\\ \scriptsize{($\pm $0.69)}\end{tabular}            &  \begin{tabular}[c]{@{}c@{}}31.15\\ \scriptsize{($\pm $0.33)}\end{tabular}             & \begin{tabular}[c]{@{}c@{}}26.11\\ \scriptsize{($\pm $0.52)}\end{tabular}                     & \begin{tabular}[c]{@{}c@{}}29.92\\ \scriptsize{($\pm $0.49)}\end{tabular}                     \\

			$\eta =$5\%        &  \begin{tabular}[c]{@{}c@{}}20.05\\ \scriptsize{($\pm $0.13)}\end{tabular}            &  \begin{tabular}[c]{@{}c@{}}24.92\\ \scriptsize{($\pm $0.10)}\end{tabular}                                & \begin{tabular}[c]{@{}c@{}}19.32\\ \scriptsize{($\pm $0.19)}\end{tabular}                      &  \begin{tabular}[c]{@{}c@{}}24.11\\ \scriptsize{($\pm $0.20)}\end{tabular}            &  \begin{tabular}[c]{@{}c@{}}19.97\\ \scriptsize{($\pm $0.11)}\end{tabular}             & \begin{tabular}[c]{@{}c@{}}26.02\\ \scriptsize{($\pm $0.06)}\end{tabular}                     & \begin{tabular}[c]{@{}c@{}}23.47\\ \scriptsize{($\pm $0.08)}\end{tabular}                      & \begin{tabular}[c]{@{}c@{}}24.59\\ \scriptsize{($\pm $0.10)}\end{tabular} \\ 
			
			\hline
	\end{tabular}}
\end{table*}

\begin{table*}[]
	\centering
	\caption{Network Overlap study of 8-baselines and RCoCo for user alignment results on TW$_B$-FS dataset in terms of $Hit@K$ (\%) and $MRR@K$ (\%), $k$=10, where standard derivations are given in brackets. The best results are in the \textbf{boldfaced}, and the second are \underline{underlined}.}
	\vspace{0.1in}
	\label{tab:t03}
	\resizebox{\linewidth}{!}{
		\begin{tabular}{c|cc|cc|cc|cc}
			\hline
			{\multirow{2}*{\diagbox[width=6.5em]{Method}{Overlap}}} & \multicolumn{2}{c|}{$\eta =$20\%} & \multicolumn{2}{c|}{$\eta =$15\%} & \multicolumn{2}{c|}{$\eta =$10\%} & \multicolumn{2}{c}{$\eta =$5\%} \\
			& $Hit@K$           & $MRR@K$          & $Hit@K$          & $MRR@K$          & $Hit@K$                & $MRR@K$                 & $Hit@K$           & $MRR@K$          \\ \hline
			
			Moana          & \begin{tabular}[c]{@{}c@{}}33.11\\ \scriptsize{($\pm $0.06)}\end{tabular}           & \begin{tabular}[c]{@{}c@{}}32.12\\ \scriptsize{($\pm $0.15)}\end{tabular}            &\begin{tabular}[c]{@{}c@{}}29.70\\ \scriptsize{($\pm $0.26)}\end{tabular}                      &\begin{tabular}[c]{@{}c@{}}28.03\\ \scriptsize{($\pm $0.21)}\end{tabular}                       & \begin{tabular}[c]{@{}c@{}}18.62\\ \scriptsize{($\pm $0.61)}\end{tabular}           & \begin{tabular}[c]{@{}c@{}}13.36\\ \scriptsize{($\pm $0.18)}\end{tabular}            &\begin{tabular}[c]{@{}c@{}}13.72\\ \scriptsize{($\pm $0.05)}\end{tabular}                      &\begin{tabular}[c]{@{}c@{}}10.05\\ \scriptsize{($\pm $0.07)}\end{tabular}                       \\

			G-CREWE        & \begin{tabular}[c]{@{}c@{}}29.13\\ \scriptsize{($\pm $0.12)}\end{tabular}            & \begin{tabular}[c]{@{}c@{}}28.20\\ \scriptsize{($\pm $0.10)}\end{tabular}             & \begin{tabular}[c]{@{}c@{}}26.37\\ \scriptsize{($\pm $0.11)}\end{tabular}                     & \begin{tabular}[c]{@{}c@{}}22.56\\ \scriptsize{($\pm $0.19)}\end{tabular}                      & \begin{tabular}[c]{@{}c@{}}20.10\\ \scriptsize{($\pm $0.39)}\end{tabular}            & \begin{tabular}[c]{@{}c@{}}15.28\\ \scriptsize{($\pm $0.26)}\end{tabular}             & \begin{tabular}[c]{@{}c@{}}16.03\\ \scriptsize{($\pm $0.10)}\end{tabular}                     & \begin{tabular}[c]{@{}c@{}}12.16\\ \scriptsize{($\pm $0.15)}\end{tabular}                     \\

			SNNA           & \begin{tabular}[c]{@{}c@{}}35.05\\ \scriptsize{($\pm $0.10)}\end{tabular}            &   \begin{tabular}[c]{@{}c@{}}33.57\\ \scriptsize{($\pm $0.12)}\end{tabular}                     &
			\begin{tabular}[c]{@{}c@{}}31.92\\ \scriptsize{($\pm $0.31)}\end{tabular}            & \begin{tabular}[c]{@{}c@{}}29.37\\ \scriptsize{($\pm $0.10)}\end{tabular}                      &  \begin{tabular}[c]{@{}c@{}}12.93\\ \scriptsize{($\pm $0.27)}\end{tabular}            &  \begin{tabular}[c]{@{}c@{}}10.29\\ \scriptsize{($\pm $0.33)}\end{tabular}             & \begin{tabular}[c]{@{}c@{}}9.89\\ \scriptsize{($\pm $0.11)}\end{tabular}                     & \begin{tabular}[c]{@{}c@{}}9.03\\ \scriptsize{($\pm $0.24)}\end{tabular}                     \\

			Meta-NA        &  \begin{tabular}[c]{@{}c@{}}31.32\\ \scriptsize{($\pm $0.25)}\end{tabular}            &  \begin{tabular}[c]{@{}c@{}}30.52\\ \scriptsize{($\pm $0.20)}\end{tabular}             & \begin{tabular}[c]{@{}c@{}}24.02\\ \scriptsize{($\pm $0.33)}\end{tabular}                     & \begin{tabular}[c]{@{}c@{}}22.15\\ \scriptsize{($\pm $0.22)}\end{tabular}                      &  \begin{tabular}[c]{@{}c@{}}13.70\\ \scriptsize{($\pm $1.67)}\end{tabular}            &  \begin{tabular}[c]{@{}c@{}}9.72\\ \scriptsize{($\pm $0.19)}\end{tabular}             & \begin{tabular}[c]{@{}c@{}}10.57\\ \scriptsize{($\pm $2.07)}\end{tabular}                     & \begin{tabular}[c]{@{}c@{}}6.16\\ \scriptsize{($\pm $0.31)}\end{tabular}                     \\

			IONE-D         &  \begin{tabular}[c]{@{}c@{}}39.25\\ \scriptsize{($\pm $0.33)}\end{tabular}            &  \begin{tabular}[c]{@{}c@{}}34.75\\ \scriptsize{($\pm $0.52)}\end{tabular}             & \begin{tabular}[c]{@{}c@{}}31.22\\ \scriptsize{($\pm $0.21)}\end{tabular}                     & \begin{tabular}[c]{@{}c@{}}25.92\\ \scriptsize{($\pm $0.39)}\end{tabular}                      &  \begin{tabular}[c]{@{}c@{}}19.55\\ \scriptsize{($\pm $0.81)}\end{tabular}            &  \begin{tabular}[c]{@{}c@{}}17.30\\ \scriptsize{($\pm $0.72)}\end{tabular}             & \begin{tabular}[c]{@{}c@{}}15.69\\ \scriptsize{($\pm $0.06)}\end{tabular}                     & \begin{tabular}[c]{@{}c@{}}11.52\\ \scriptsize{($\pm $0.67)}\end{tabular}                     \\

			NeXtAlign      &  \begin{tabular}[c]{@{}c@{}}\underline{45.10}\\ \scriptsize{($\pm $0.20)}\end{tabular}           &  \begin{tabular}[c]{@{}c@{}}{37.06}\\ \scriptsize{($\pm $0.33)}\end{tabular}            & \begin{tabular}[c]{@{}c@{}}{33.91}\\ \scriptsize{($\pm $0.14)}\end{tabular}                     & \begin{tabular}[c]{@{}c@{}}\underline{30.09}\\ \scriptsize{($\pm $0.31)}\end{tabular}                      &  \begin{tabular}[c]{@{}c@{}}{25.24}\\ \scriptsize{($\pm $1.51)}\end{tabular}            &  \begin{tabular}[c]{@{}c@{}}\underline{21.96}\\ \scriptsize{($\pm $0.52)}\end{tabular}             & \begin{tabular}[c]{@{}c@{}}{13.20}\\ \scriptsize{($\pm $1.12)}\end{tabular}                     & \begin{tabular}[c]{@{}c@{}}9.37\\ \scriptsize{($\pm $0.48)}\end{tabular}                     \\

			NAME           &  \begin{tabular}[c]{@{}c@{}}40.67\\ \scriptsize{($\pm $0.18)}\end{tabular}            &  \begin{tabular}[c]{@{}c@{}}\underline{37.15}\\ \scriptsize{($\pm $0.12)}\end{tabular}             & \begin{tabular}[c]{@{}c@{}}\underline{36.20}\\ \scriptsize{($\pm $0.23)}\end{tabular}                     & \begin{tabular}[c]{@{}c@{}}29.53\\ \scriptsize{($\pm $0.25)}\end{tabular}                      &  \begin{tabular}[c]{@{}c@{}}21.13\\ \scriptsize{($\pm $0.92)}\end{tabular}            &  \begin{tabular}[c]{@{}c@{}}17.93\\ \scriptsize{($\pm $0.43)}\end{tabular}            & \begin{tabular}[c]{@{}c@{}}14.35\\ \scriptsize{($\pm $0.39)}\end{tabular}                     & \begin{tabular}[c]{@{}c@{}}{11.03}\\ \scriptsize{($\pm $0.33)}\end{tabular}                     \\

			HUIL           &  \begin{tabular}[c]{@{}c@{}}38.55\\ \scriptsize{($\pm $0.20)}\end{tabular}            &  \begin{tabular}[c]{@{}c@{}}36.33\\ \scriptsize{($\pm $0.30)}\end{tabular}             & \begin{tabular}[c]{@{}c@{}}33.71\\ \scriptsize{($\pm $0.12)}\end{tabular}                     & \begin{tabular}[c]{@{}c@{}}28.25\\ \scriptsize{($\pm $0.36)}\end{tabular}                      &  \begin{tabular}[c]{@{}c@{}}\underline{26.67}\\ \scriptsize{($\pm $0.56)}\end{tabular}            &  \begin{tabular}[c]{@{}c@{}}21.11\\ \scriptsize{($\pm $0.29)}\end{tabular}            & \begin{tabular}[c]{@{}c@{}}\underline{20.16}\\ \scriptsize{($\pm $0.12)}\end{tabular}                     & \begin{tabular}[c]{@{}c@{}}\underline{14.22}\\ \scriptsize{($\pm $0.26)}\end{tabular}                     \\ \hline

			\textbf{RCoCo} & \begin{tabular}[c]{@{}c@{}}\textbf{47.93}\\ \scriptsize{{($\pm $0.15)}}\end{tabular} & \begin{tabular}[c]{@{}c@{}}\textbf{39.30}\\ \scriptsize{{($\pm $0.25)}}\end{tabular} & \begin{tabular}[c]{@{}c@{}}\textbf{43.35}\\ \scriptsize{{($\pm $0.39)}}\end{tabular}  & \begin{tabular}[c]{@{}c@{}}\textbf{31.82}\\ \scriptsize{{($\pm $0.15)}}\end{tabular}  & \begin{tabular}[c]{@{}c@{}}\textbf{37.20}\\ \scriptsize{{($\pm $1.03)}}\end{tabular}  & \begin{tabular}[c]{@{}c@{}}\textbf{23.05}\\ \scriptsize{{($\pm $0.35)}}\end{tabular}  & \begin{tabular}[c]{@{}c@{}}\textbf{27.08}\\ \scriptsize{{($\pm $0.03)}}\end{tabular}  & \begin{tabular}[c]{@{}c@{}}\textbf{19.32}\\ \scriptsize{{($\pm $0.19)}}\end{tabular}  \\ \hline
	\end{tabular}}
\end{table*}

\subsection{Discussion on Representation Dimension}
In the section, we study the dimension of user representation, and discuss the expressiveness of different manifold in the meanwhile.

\begin{figure*}[htbp]
	\centering
	\includegraphics[width=13cm,height=5cm]{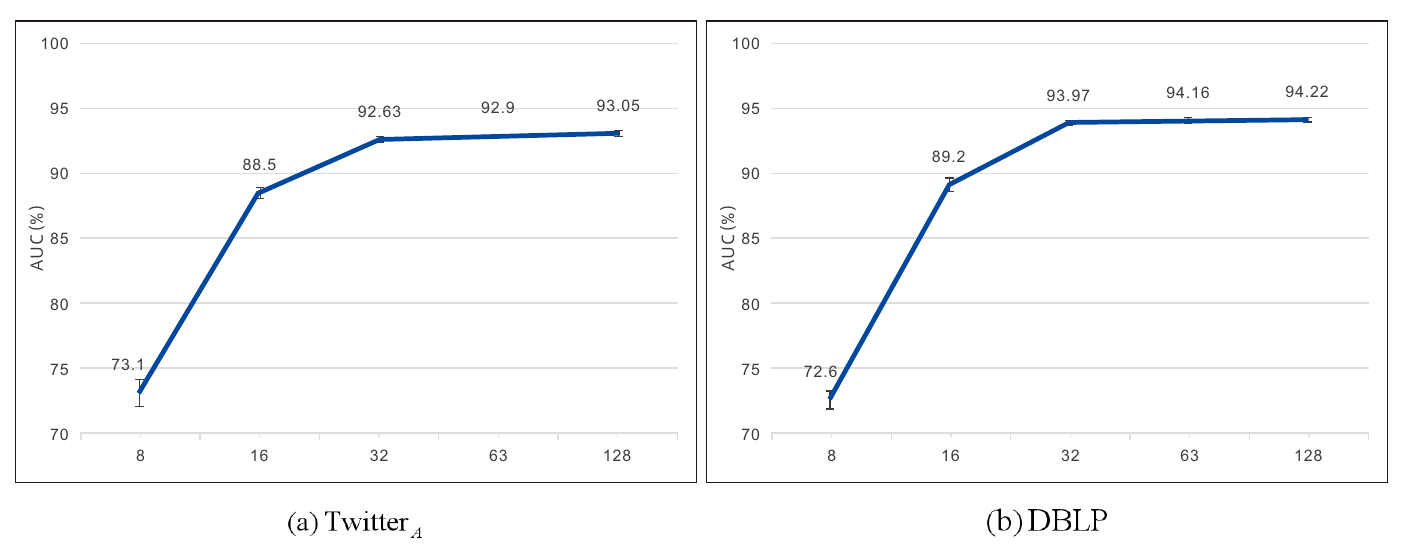}
	\caption{Results of RCoCo performs intra-link prediction on Twitter$_A$ and DBLP datasets in different dimensions $\{8, 16, 32, 64, 128\}$ (in terms of AUC).}
	\label{Fig.4}
\end{figure*}

\begin{figure*}[htbp]
	\centering
	\includegraphics[width=13cm,height=5cm]{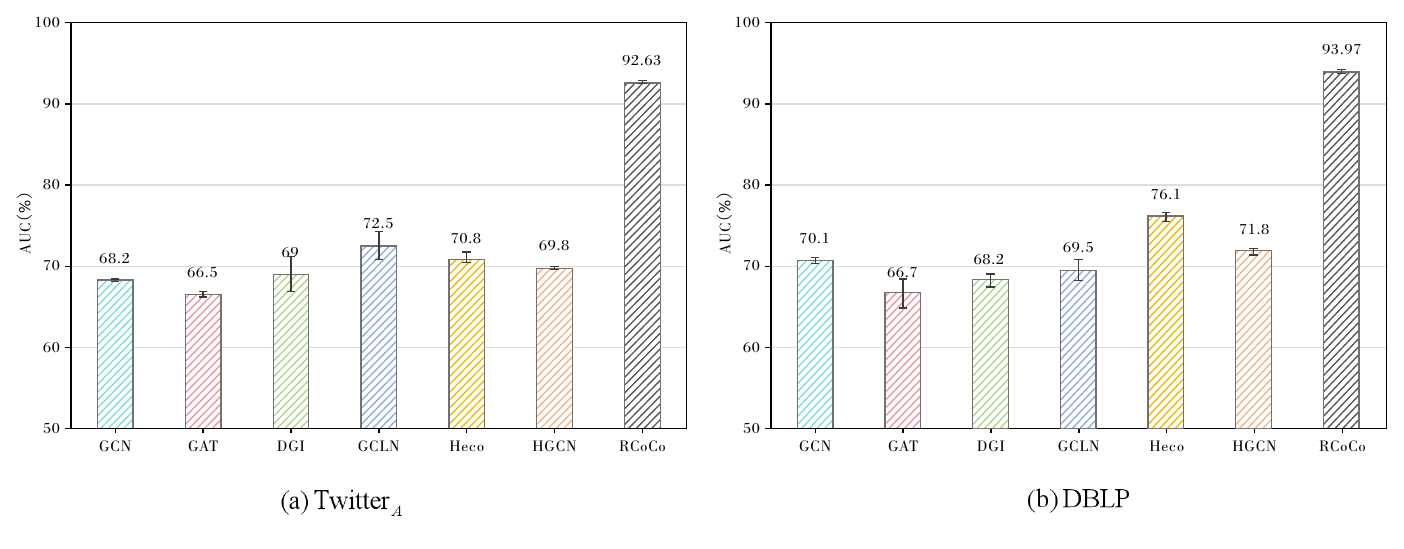}
	\caption{Results of Euclidean baselines intra-link prediction on Twitter$_A$ and DBLP datasets in dimension 16 (in terms of AUC).}
	\label{Fig.5}
\end{figure*}

We vary the representation dimension of RCoCo in $\{8, 16, 32, 64, 128\}$. We do intra-link prediction via RCoCo on Twitter$_A$ and DBLP datasets, and report the empirical results of  under different dimensions (in terms of AUC) in Fig. \ref{Fig.4} (a) and (b), respectively.
It is interesting to find that RCoCo  is able to achieve satisfactory performance with low-dimensional embeddings (e.g., $d=16$ as shown in Fig. \ref{Fig.4}), and does not receive significant AUC gain when the representation dimension further increases.
In addition, we investigate in the low-dimensional embeddings of Euclidean baselines, and summarize the prediction results in Fig. \ref{Fig.5}, where the dimension is set as $16$.
Unlike the Riemannian counterpart, Euclidean baselines lacks the ability to make promising prediction with low-dimensional embeddings. Thus, Euclidean baselines are usually designed with $256$-dimensional embeddings or even higher dimensions.
The reason lies in the expressiveness of the manifold itself. The curved manifolds\footnote{Curved manifolds refers to hyperbolic and hyperspherical spaces as they are negatively curved and positively curved, respectively.} typically own better expressiveness than the flat one (Euclidean space). 
For instance, a $2$-dimensional hyperbolic space is capable to embed any graph structure with arbitrary low distortion. Note that, the distortion is the metric to evaluate the wellness of an embedding.
If an embedding embeds a node $v_i$ in the graph to a point $\mathbf x_i$ in the manifold, the distortion is formally defined as follows,
\begin{equation}
	embedding: \mathcal V \to \mathcal M, \quad distortion=\frac{1}{N^2} \sum_{ij}\left| \frac{d_G(v_i, v_j)}{d_\kappa(\mathbf x_i, \mathbf x_j)}-1\right|,
\end{equation}
where $d_G$ and $d_\kappa$ denote the shortest distance in the graph and the distance in the manifold, respectively.
The lower the distortion, the better the embedding.
In contract, the distortion is still not bounded under $128$-dimensional embeddings in Euclidean space.
In fact, for a given dimension, the curved manifolds have much larger volume for representation compared to the Euclid.
Accordingly, for a given graph structure, embeddings in the curved manifolds tend to achieve competitive (or even better) expressiveness with lower dimensions in the Euclidean space.
Further exploration on dimension selection is out of the scope of this paper. 

\section{Related Work}

\subsection{Intra- \& Inter-network Link Prediction}

\subsubsection{Link Prediction}
The existing link prediction methods can be classified into two types: Graph-Embedding and Subgraph-Based methods.

\noindent \textbf{Graph-Embedding.} 
Early link prediction methods usually use graph embedding, such as DeepWalk \cite{kddDeepWalk14} and node2vec {\cite{kddnode2vec16}}.
These methods are embedded in Euclidean space, and they are typically regardless of the inherent geometry of real social networks. In recent years, some researchers have proposed the superiority of Non-Euclidean spaces for graph embedding {\cite{nipsNicke17}}, {\cite{tkddZhang22}}. For example, Wang et al. \cite{{IEEEWang20}}, Sun et al. {\cite{DBLP:conf/icdm/0008Z0WDSY20}}, Bai et al. {\cite{DBLP:conf/www/BaiNZZY23}} propose to embed node and/or community in hyperbolic space.
However, given the complexity of the internal relationships within social networks, it is challenging to  directly determine whether embedding graph in a Euclidean space or a hyperbolic space is appropriate.

\noindent \textbf{Subgraph-Based.}
The subgraph-based approach involves learning features from subgraphs to predict connections. 
Fang et al. \cite{TKDEFang23} proposed an architecture for link prediction called Neural Networks with Elementary Subgraphs Features (NNESF), which has relatively low computational complexity and a small number of hyperparameters. 
Lai et al. proposed the ARCLink model \cite{CIKMLai21}, which is capable of creating a more efficient subgraph vector representation. This representation enables hierarchical aggregation of node features based on the learned node importance.
Jiao et al. {\cite{cikmJiao19}} introduces a hierarchical graph attention mechanism.
However, subgraph-based methods are significantly influenced by nodes, making it essential to improve the importance quantification of nodes in the graph relative to the target nodes. 

Most of the above link prediction methods are represented in Euclidean space. However, different social networks often have different geometric shapes, and the representation spaces of different layers are not the same in the multiplex network. Thus, our proposed RCoCo considers the inherent geometry of each layer of social network, and addresses the link prediction problem via generic constant curvature space.

\subsubsection{User Alignment}
User alignment aims to learn a matching between the same entities across multiple network.
Existing network alignment methods can be divided into two categories including Semi-Supervised and Unsupervised Methods.  

\noindent \textbf{Semi-Supervised.} These methods {\cite{infocom18}}, {\cite{ijcaiLiu16}}, \cite{tkdeSunZWJWSY23} use known anchor links or other constraints to complete user alignment. Kong et al. proposed a method named MNA {\cite{cikmKongZY13}}, which can extract features from multiple heterogeneous networks for anchor link prediction. Liu et.al {\cite{ijcaiLiu16}} used follower/followee relationship to learn alignment-oriented network embedding. 
\cite{tkdeSunZWJWSY23} introduce an optimization framework over GNNs to conduct alignment. 
\cite{SunL23tweb} focus on the group-level alignment, which is orthogonal to our study in this paper.
Semi-supervised methods typically rely on the adequate anchor nodes, which limits the wide use in real cases.  


\noindent \textbf{Unsupervised.}
Unsupervised methods highly  rely on discriminative feature extraction {\cite{icdeHuynh20}}, {\cite{SunL23tist}}, {\cite{kddGao21}}. 
\cite{SunL23tist} explore the matrix cones for informative user embeddings.
Recently, 
Zhou et al. {\cite{TKDDZhOU22}} propose an alignment method based on the unsupervised adversarial learning.
This approach constructs two cross-network alignment translation models to train unsupervised alignment, without  prior alignment information.


The behaviors of users are influenced not only by a single social network but also by the different networks they are part of. Existing studies focus on independently analyzing intra-network link prediction and inter-network user alignment. However, our work aims to enhance link prediction by considering the impact of both intra-network and inter-network aspects together.

\subsection{Graph Contrastive Learning}

Graph contrastive learning explores the similarities from the graph itself, and contrast positive and negative pairs to learn graph representations, boosting the performance of GNNs {\cite{niu2020gmta}}.
So far, there is still a lack of research on graph contrastive learning for collective link prediction,  although it has been widely used in various fields.
We briefly review two main types of graph contrastive learning: Unsupervised  and Self-Supervised Graph Contrastive Learning.

\noindent \textbf{Unsupervised Graph Contrastive Learning}.
S3-CL {\cite{aaaiDing023}} method is a neural network with structural and semantic contrastive learning, which addresses the limitations of capturing graph knowledge with similar feature nodes.
The GCLN {\cite{IEEEWU22}} explores the interaction of attraction and repulsion on the graphs. The attraction encourages features from both graph domains to be consistent, and the repulsion ensures the differentiation of features, devoted to resolve the limitations of cross-graph domains.

\noindent \textbf{Self-Supervised Graph Contrastive Learning}.
The HeCo {\cite{tkdeLiu23}} method is a novel self-supervised learning based on the co-contrast learning mechanism. It captures both local and higher-order structures via contrasting the network-mode view and a meta-path view.
CGMN {\cite{ijcaiJin22}} is based on self-supervised graph similarity calculation, which identifies the differences in nodes across different graphs.
Wu et al. {\cite{tkdeWu23}} discuss the comparison of graphs at the same scale and across different scales.


Unlike the contrastive learning in vision domain, the issue of  augmentations largely remains open in graph contrastive learning. 
In this paper, we  propose a novel self-augmentation approach based on exploring the user and community views of the social network. 
In this way, we improve the accuracy of link prediction in multiplex network while minimizing reliance on tagged anchor users.

\subsection{Riemannian Graph Learning}

In the past decade, learning on graphs is conduced  in Euclidean spaces, e.g., network embedding and GNNs. 
The early practices, such as node2vec {\cite{kddnode2vec16}} and DeepWalk {\cite{kddDeepWalk14}}, embed nodes into a low-dimensional Euclidean space. 
Riemannian space (e.g., hyperbolic and other manifold) provides an exciting alternative. 
In recent years, researchers  show  that hyperbolic spaces are more suitable for embedding graphs with tree-like structures than Euclidean spaces, and have proposed GNNs in hyperbolic space.
As one of the early representatives, HGCN \cite{nipsChamiYRL19} maps Euclidean node features to hyperbolic spaces, and defines a convolutional graph networks in hyperbolic spaces. 
Subsequently, Bai et al. \cite{DBLP:conf/www/BaiNZZY23}, Wang et al. \cite{IEEEWang20}, and Liu et al. \cite{apinLiu23} have also conducted learning on graphs in hyperbolic space. 
Beyond the tree-likeness of the structures, $\kappa$-GCN \cite{icmlcurvatureGCN20} extends the Euclidean graph convolutional network to $\kappa$-stereomodels with arbitrary curvature.
Recent studies \cite{aaaiFan21,HVGNN,aaai22SelfMix} also explore the tangent space of Riemannian space to study graph learning. 
Though Euclidean and Riemannian GNNs have a collection of parallel notions, but it is noteworthy to mention that the computational tools in Euclidean space cannot be directly applied to Riemannian manifolds.

Motivated by the superior expressiveness in Riemannian space, Riemannian GNNs has been successfully applied to a wide range of applications (such as recommender systems), but they has not yet been introduced  to the problem of collective link prediction across multiplex network, to the best of our knowledge.
In addition, we further consider the inherent geometry of each layer of multiplex network.



\section{Conclusion}
In this paper, we introduce a novel method for contrastive Collective Link Prediction across Multiplex Network in Riemannian Space: RCoCo, and predict intra- and inter-links simultaneously. Specifically, we design a curvature estimator to adapt the geometry of each network, and then we construct a curvature-aware graph attention network ($\kappa-$ GAT) to conduct intra- and inter-network attention aggregation in the manifold with the curvature estimated. Thereafter, we conduct contrastive learning in the manifold via loss functions. For intra-contrastive learning, we group the community members as a supernode with I-Louvain algorithm, and contrast the original node view and generated supernode view. For inter-contrastive learning, we maxmize the representation agreement between anchor user in the common tangent space. Extensive experiments show the superiority of RCoCo.

\section*{Acknowledgments}
The authors of this paper were supported in part by National Natural Science Foundation of China under grant 62202164 and 62102273.
Corresponding Authors: Li Sun and Yong Yang.

\bibliographystyle{unsrt}
\bibliography{reference}

\end{document}